\documentclass [12pt]{article}
\usepackage {amssymb}
\usepackage {amsmath}
\usepackage [cp1251]{inputenc}
\usepackage{graphicx}
\usepackage{cite}
\usepackage {longtable}

\title{Cosmology based on entropy}
\author{$^{1}$\textbf{Yu.L.Bolotin}, $^{2,3}$\textbf{V. V. Yanovsky}}
\begin{document}
\maketitle
$^{1}$\textit{National Science Center ''Kharkov Institute of Physics and Technology'', 1 Akademicheskaya str.,  Kharkov, 61108, Ukraine}

$^{2}$\textit{V. N. Karazin Kharkiv National University, 4 Svobody Sq., Kharkov, 61022, Ukraine}

$^{3}$ \textit{Institute for Single Crystals, NAS Ukraine, 60 Nauky Ave., Kharkov, 61001, Ukraine}

\abstract{At present, there is practically no doubt that general relativity is closely related to gravity. Moreover, after the work of Jacobson, Padmanabhan  and others, it became clear that a thermodynamic interpretation of Einstein's relativistic equations is possible. On the other hand, we are witnessing the conceptual problems of the SCM (the problem of the cosmological constant, the problem of coincidences) and many years of futile attempts to directly fix the main components of the model (dark energy and dark matter). The combination of these two factors gave rise to a natural desire, at least at the phenomenological level, to build a cosmological model that represents the synthesis of gravity and thermodynamics and does not include components of an unknown nature. It is this model -- entropic cosmology -- hat is considered in this review. We have set as our goal, omitting the details that can be found in the references given, to present the conceptual foundations of the model.

The key role of the model is played by the thermodynamics of the horizon, or, more specifically, its thermodynamic characteristics -- entropy and temperature. In entropic cosmology, the apparent horizon of the Universe is usually chosen as the horizon.  For a spatially flat Universe (such a Universe is considered in this review) coincides with the Hubble horizon. We present arguments justifying this choice. Much attention is paid to the problem of choosing the entropy of the cosmological horizon. By calculating the energy flux across through the Hubble horizon, we demonstrate that the fulfillment of the Clausius relation and the standard Friedmann equations (i.e. GR) uniquely lead to the Bekenstein entropy. Due to the importance of this result, we also reproduced it within the framework of emergent cosmology.

Two formulations of the basic equations of entropic cosmology are considered in detail. The first of them is based on the direct consideration of entropic forces (or, equivalently, the consideration of the contribution of surface terms when varying the Hilbert-Einstein action). Contribution of entropic forces is achieved by modification of the Friedmann equation and the accelerator equation by introducing phenomenological derive terms. Two types of modification are considered: for the non-dissipative Universe ($\Lambda (t)$-model) and for the dissipative Universe (BV-model).

An alternative approach to formulation of the basic equations of entropic cosmology is to treat the contribution of surface terms as holographic (entropic) dark energy. This allows us to keep the standard form of Friedmann equations. The equation of state and other characteristics of this type of dark energy depend on the choice of entropy on which it is based. As an example, we considered entropy dark energy constructed using Barrow's entropy.
The review considers the foundations of a fundamentally new approach to the evolution of the Universe proposed by Padmanabhan, in which space-time itself is an emergent  structure}.

\tableofcontents

\section{Introduction}
\label{p1}

According to the current cosmological paradigm, our Universe is in accelerating expansion at present \cite{1s,2s}.   However, the physical origin of the observed cosmic acceleration still remains the greatest mystery. Universe filled with ''ordinary'' components (matter and radiation) should eventually slow down the expansion.  Any attempt to explain the observed accelerated expansion of the Universe, while remaining within the framework of the general theory of relativity (GR), can be made either by modifying the field equations, or by including components of an unknown nature in the energy-momentum tensor.  At the phenomenological level, the introduction of dark components can be avoided by modifying the Friedmann equations, i.e. changing the nature of t evolution of ordinary components (matter and radiation) in the required direction. It turned out that the required goal can be achieved in various ways, including: replacing an ideal fluid with a viscous one, introducing sources into the conservation equation, transforming the equations of state, and many other ways.
Among the models that do not require the introduction of new components, a special place is occupied by the scenario of the evolution of the Universe, called entropic cosmology\cite{3s,4s}. As the name implies, the considered scenario of cosmological evolution is based on two fundamental concepts of physics - entropy (thermodynamics) and cosmology (gravity).

Einstein used geometry to explain the gravity. Geometry is one of the most general ideas born by the science. It is not surprising that the axiomatics of geometry has become a model for a number of fundamental physical theories. So, in particular, classical thermodynamics is built on an axiomatic basis. This axiomatic theory has an extremely general character and a high degree of universality.Not surprisingly, there is a natural desire to expand the range of problems for which thermodynamics can be used.

The intriguing connection between gravity and thermodynamics came into focus in the last quarter of the twentieth century. At present, the existence of a close relationship between gravity and thermodynamics is widely recognized. Such a connection was first demonstrated in the context of black hole physics\cite{5s,6s,7s,8s,9s}.  Within the framework of the semiclassical description, it was shown that black holes behave like a black body with radiation, called Hawking radiation\cite{10s,11s}. The radiation temperature is proportional to surface gravity. Further progress in this direction is associated with the work of Jacobson\cite{12s}, who obtained the field equations of general relativity from the Clausius relation on the local Rindler causal horizon\cite{13s,14s} with an entropy proportional to the area of the horizon.

It soon became clear that the concept of the horizon is not the prerogative of black holes\cite{15s}. Therefore, it seems natural that the thermodynamics of black hole horizons has been extended to a wider range of objects. This circle includes: black hole horizons, horizons arising within the framework of general relativity, cosmological horizons, arbitrary surfaces of space-time, screens in holographic dynamics.
In particular, shortly after Hawking's discovery of the temperature associated with the black hole horizon, it became clear that this result is not limited to black holes only: in any space-time with a horizon, the latter can be assigned a temperature. Thus, an observer moving at an accelerated rate in vacuum perceives the horizon as an object with a temperature proportional to its acceleration\cite{16s}.

The situation with entropy is much more complicated. Is it possible to attribute entropy to any horizon? We can say that the inclusion of horizons other than the event horizon of a black hole gives rise to a number of problems. These problems can be divided into two groups. The first includes problems inherited from the thermodynamics of black holes. Apparently, the most important problem of this group is the search for a logically impeccable answer to the question: does gravitational dynamics formulated, for example, in the form of GR equations, have any relation to thermodynamics in general and to horizon thermodynamics in particular? A wide range of answers to this question, from the direct derivation of the GR equations\cite{23s} within the framework of thermodynamics to depriving gravity of the status of a fundamental force, indicates the absence of a generally accepted point of view on the problem. Among the problems of the second group, we note that many successful attempts to interpret the entropy of black holes cannot be implemented to interpret the entropy of other horizons. Thus, a unified understanding of horizon thermodynamics, covering both entropy and temperature, has not yet been achieved.

\section{Horizons, entropy, temperature}

The concept of the cosmological horizon is a fundamental concept of entropic cosmology. Let's start with a discussion of what is meant in this context by the concept of horizon in general and the cosmological horizon in particular. Before the appearance of the classic article by Rindler \cite{14s}, many definitions of the concepts ''horizon'' led to confusion.

The most physically transparent is the so-called particle horizon, in the simplest case a spherical surface that limits the region of the Universe available for observation by an comoving observer at a present time. Its radius as a function of the scale factor $a$
\begin{equation}\label{eh1}
  l_p (a)= a \int_{t_i}^{t}\frac{dt}{a(t)}=a \int_{0}^{a}\frac{d \tilde{a}}{\tilde{a}^2 H(\tilde{a})}
\end{equation}
where  $t_i$ - is the initial moment of time when the scale factor vanishes. The concept of a particle horizon is useful for decelerated expansion of the Universe (domination of radiation or matter) for which  $l_p$ is finite. If, however, the Universe is in the de Sitter phase (accelerated expansion), there is no particle horizon, since  $l_p = \infty $. In this case we can also consider the additional concept of the event horizon, which is the boundary of the region of space that a commoving observer will ever see, namely
\begin{equation}\label{eh2}
  l_{event} (a)= a \int_{t}^{\infty}\frac{dt}{a(t)}=a  \int_{a}^{\infty}\frac{d \tilde{a}}{\tilde{a}^2 H(\tilde{a})}
\end{equation}
This integral diverges for flat and open FLRW universes without a cosmological constant, and hence there is no event horizon for them (all future events are eventually available). On the contrary, for LCDM (SCM) the integral is finite. If, $a \gg 1$  then
\begin{equation}\label{eh3}
  l_{event} (a) \simeq \sqrt{\frac{3}{\Lambda}}=\frac{1}{H}
\end{equation}
where $H$ - is the asymptotic value of the Hubble parameter. Thus, for accelerated expanding universes  there is event horizon, but no particle horizon. The above two definitions of the horizon are obviously different concepts, the former referring to our knowledge of events in the past, while the latter refers to events in our future.

Let us now introduce the extremely efficient and physically transparent concept of a "Hubble sphere", a sphere with radius  $R_H =H^{-1}$. Galaxies inside the Hubble sphere are moving away from us at a speed less than the speed of light, and outside --- at a speed greater than the speed of light. The Hubble sphere does not coincide with the observable universe. The observable universe is bounded by a particle horizon. In a static Universe  $R_H =H^{-1}= \infty$, while  radius of  the particle  horizon is finite for the finite lifetime of the Universe. In what follows, we will refer to the Hubble sphere as the Hubble horizon.

To discuss the thermodynamics of the expanding Universe, the concept of cosmological apparent  horizon, which differs from the horizons described above, is useful. This definition is more appropriate than the event horizon, since the latter requires knowledge of all future evolution and the causal structure of space-time. The apparent horizon also has the advantage that, when used as a holographic screen, the laws of classical dynamics are reproduced.

Sacrificing generality, we give the definition of the apparent horizon for a homogeneous and isotropic FLRW Universe whose metric can be represented as
\begin{equation}\label{ehd1}
  ds^{2}=h_{ab}dx^{a}dx^{b}+ \tilde{r}^{2}d\Omega^{2}_{n-1}
\end{equation}
where $ \tilde{r}=a(t)r$, $x^{0}=t$, $x^{1}=r$, and  $h_{ab} =diag (-1, \frac{a^{2}}{1-kr^{2}})$,  $\Omega_{n}=\frac{\pi^{n/2}}{\Gamma(\frac{n}{2} +1)}$ is two-dimensional tensor. The radius of the apparent horizon is determined by the equation\cite{17s,18s,19s}.
\begin{equation}\label{ehd2}
  h_{ab}\partial_{a}\tilde{r}\partial_{b}\tilde{r}=0
\end{equation}
Solving this equation, we find the radius of the apparent horizon
\begin{equation}\label{ehd3}
  \tilde{r}_{A}=\frac{1}{\sqrt{H^{2}+k/a^{2}}}
\end{equation}
It is easy to see that for a flat de Sitter Universe, the radius of the apparent horizon, the Hubble horizon, and the cosmological event horizon have the same constant value $H^{-1}$. Note that although cosmological horizons (particles or events) exit for a limited number of options for the evolution of the  FLRW Universe the apparent horizon and the Hubble horizon always exist.
The concept of the horizon naturally arises when trying to estimate the entropy of  Universe. There are two approaches to quantify the entropy of  Universe\cite{20s}.

In the first of these approaches, we track the entropy of some sufficiently large commoving volume bounded by a closed (commoving) surface. Such a system is effectively isolated because the large scale homogeneity and isotropy of the universe imply no flow of entropy into or out of the commoving volume. The total entropy of a commoving volume is the sum of the entropies of the objects within that volume. The assumption of large-scale homogeneity makes it possible to track the entropy of matter beyond a chosen commoving volume. A reasonable choice of commoving volume in this scheme is the commoving sphere, which currently corresponds to the observable universe. The entropy of the event horizon of the universe is absent in such a scheme.

In the second approach, the considered system is limited by some model-dependent cosmological horizon instead of the comoving surface boundary. In this case, the migration of matter across the horizon cannot be neglected, and the time-dependent entropy of the horizon must be included in the budget.

The two approaches described briefly above can be formulated as follows\cite{20s}:
\begin{enumerate}
  \item The total entropy in a sufficiently large comoving volume of the universe does not decrease with cosmic time, \[dS_{comuving \,\, volume} \geq 0\]
  \item The total entropy of matter contained within the cosmic event horizon (CEH) plus the entropy of the CEH itself, \[ dS_{CEH}+ S_{CEH \,\, inter}\geq 0\]
\end{enumerate}
In this work, we use the second approach, assuming that total entropy
\begin{equation}\label{eh4}
  S_{total}=S_h +S_m
\end{equation}
Let us now turn to the thermodynamic characteristics of (cosmological) horizons. The introduction of such characteristics became possible after establishing a connection between the dynamics of a black hole (including the dynamics of the horizon), controlled by gravity, and the laws of thermodynamics\cite{9s}.

The introduction of the entropy of a black hole turned out to be necessary to resolve the paradox formulated by A. Wheeler: an external observer can drop material with non-zero entropy into an area beyond the horizon that is inaccessible to observation, thereby reducing the entropy accessible to external observers. This conceivable experiment clearly contradicts the second law of thermodynamics. In an attempt to resolve this paradox, Bekenstein\cite{6s,7s} came to the conclusion that the horizon of a black hole should be assigned an entropy proportional to its area $A$
\begin{equation}\label{eh5}
  S=\frac{A}{4G}
\end{equation}
Of course, Bekenstein knew the theorem proved by Hawking: in any classical process involving black holes, the sum of the areas of black hole horizons cannot decrease. The theorem is a direct analogy with the behavior of entropy in classical thermodynamics. This helped him understand the key role of the surface in the thermodynamics of a black hole.

By ascribing entropy to the horizon of a black hole, we run into a new problem. Any object with energy and entropy must have a non-zero temperature and, as a result, must radiate. This statement, at first glance, contradicts the very definition of a black hole. However, the discovery of Hawking radiation with a Planck spectrum\cite{10s,11s} showed that more than just entropy can be consistently attributed to the horizon of a black hole but also temperature (Hawking temperature),
\begin{equation}\label{eh6}
  T =\frac{\hbar c^3}{8 \pi G k_B M}
\end{equation}
The assumption about the existence of the temperature of the horizon is equivalent to the statement about the existence of its microstructure. It is important to note that this conceptual statement does not need any experimental evidence. By assigning a temperature to the horizon, we postulate the possibility of heating space-time in the same way that a solid body or gas can be heated. Hawking radiation can be used as a low-power heating device.  In other words, the temperature of the horizon is as real as the ordinary temperature of a macroscopic body. The same result can be achieved by bringing the heated object into accelerated motion relative to vacuum. Temperature experienced by a uniformly accelerating detector in a vacuum (Unruh temperature\cite{16s}) is given by
\begin{equation}\label{ehd4}
  T_U =\frac{1}{2\pi k_B}\frac{\hbar}{c}a
\end{equation}
where $a$ is detector acceleration

The reality of the horizon temperature means that space-time has a microstructure. In the case of a solid or a gas, we know the nature of this microstructure. This makes it possible to construct thermodynamics on the basis of statistical mechanics. This is impossible when constructing the thermodynamics of the horizon (and in a broader sense of the thermodynamics of space-time). At present, we do not know the microscopic degrees of freedom of space-time. However, thermodynamics - in contrast to statistical mechanics is insensitive to the details of the microstructure and macroscopic description can be constructed without information about hides microscopic degrees of freedom.

\section{Entropic force}

In modern physics, two opposite tendencies are clearly traced. On the one hand, this is a search for interactions outside the Standard Model for solving microphysics problems. On the other hand, a return to the traditional methods of classical physics at a new level to solve the newly arisen problems of macrophysics. One of these possibilities of classical thermodynamics, which does not require the introduction of new forces or exotic dark components, will be considered in this section.
Let's consider two vessels connected to each other, one of which is filled with gas. If the partition separating them is opened, then the relaxation process will begin, leading to the equalization of the gas density in both connected vessels. What force causes the densities to equalize? At first glance, the answer is obvious: relaxation to the equilibrium state is provided by gas pressure. This answer immediately raises the next question. What is the nature of pressure force?

The answer to this question is given by classical thermodynamics. It turns out that the number of states of a gas with a fixed macroscopic energy, in which the gas is located in one of the vessels, is incalculably less than the number of states in which it is uniformly distributed over both vessels. Nature chooses the most probable state.

Note that whatever the nature of this force, it is not associated with any fields that would serve as carriers of this force. The cause of the force is purely statistical. Hence the natural definition of these forces must be given in thermodynamic terms such as entropy and temperature. We will henceforth call these forces entropic forces.

An entropic force is an effective macroscopic force that arises in a systems with many degrees of freedom due to the statistical tendency to increase entropy. The absolute value of the force is determined by the difference between the entropies of the initial and final states and does not depend on the details of microscopic dynamics. It is important to note that there is no fundamental field associated with the entropic forces.
Entropic forces usually arise in macroscopic systems such as colloids. Biophysics also provides many examples of entropic forces. Thus, large colloid molecules suspended in a thermal medium of smaller particles experience additional entropic pressure. Osmosis is another phenomenon caused by the entropic force.

Perhaps the best-known physically transparent example of entropic forces is the elasticity of a polymer molecule. A polymer molecule can be modeled by connecting many monomers of a fixed length, where each monomer is free to rotate around its attachment points. Each of these configurations has the same energy. When a polymer molecule is immersed in a thermal bath, it prefers to take on random, intricate configurations, as they are entropy favored. The number of configurations of an entangled polymer far outnumbers the number of configurations that can be realized in a stretched polymer. The statistical tendency of the transition to the state of maximum entropy is transformed into a macroscopic force, in this case, the force of elasticity.

Let us calculate the entropic force acting on the Hubble horizon, assuming that it has the Bekenstein entropy and the Hawking temperature. The logic for calculating the entropy force is as follows\cite{21s}:
\begin{enumerate}
  \item We attribute Hawking temperature to the horizon.
  \item There is a heat flux across the horizon  $\Delta Q =T \Delta S$
  \item This heat flow generates an entropic force \[F_e \Delta x =T dS\]
\end{enumerate}
For the Bekenstein entropy $S=\frac{A}{4 G}$ we find the entropic force and entropic pressure
\[F_e =-\frac{\Delta Q}{\Delta r}=- T\frac{\Delta S}{\Delta R}= -\frac{\hbar}{k_B}\frac{H k_B c^3}{2\pi G \hbar} \left( \frac{c}{H} \right) = - \frac{c^4}{G} \]
\begin{equation}\label{eh7}
  p_e =\frac{F_e}{A}=-\frac{c^2}{4 \pi G}H^2
\end{equation}
In a certain sense, this force and pressure correspond to the maximum achievable values of these quantities in our Universe\cite{22s}.

\section{Entropic cosmology paradigm}

Entropic cosmology is an attempt to build a model of the Universe without resorting to the introduction of dark energy and without going beyond the canonical general relativity. The main structural elements of entropic cosmology are general relativity, thermodynamics of black holes, the holographic principle, and entropic forces.

Entropic cosmology was first formulated in\cite{3s,4s}. In these works, the authors presented a phenomenological model that includes surface terms that are usually ignored in GR. They showed that this model leads to an accelerated expansion of the Universe without the inclusion of dark energy. It is important to note that if we include entropic forces in our consideration, then the fact of the accelerated expansion of the Universe does not depend on the specific choice of the cosmological model  (within the framework the entropic cosmology).

Gravitational action functionals in a wide class of theories have volume and surface terms. Therefore, we will schematically represent the Hilbert-Einstein action, including the contribution of matter as the sum of the volume and surface integrals
\begin{equation}\label{eh8}
  I= \int_M (R+L_m) + \frac{1}{8 \pi} \oint K
\end{equation}
Here $R$ is the scalar curvature,  $L_m$ is the Lagrangian of matter filling the Universe, and $K$is the surface curvature of the boundary.  Applying variation procedures to this action gives the usual Einstein GR equations with the addition of a contribution from surface terms:
\begin{equation}\label{eh9}
  R_{\mu \nu}- \frac{1}{2} R g_{\mu \nu}=8 \pi G T_{\mu \nu}+ \text{surface \quad terms}
\end{equation}
In the traditional approach, we completely ignore the surface term (or cancel it with a counter term) and get the field equation from the volume term in action. Therefore, any solution of the field equation obtained by such a procedure is logically independent of the nature of the surface term. The situation changes when the surface term contribution is estimated at the horizon. It is he who generates the thermodynamic characteristics of the horizon. This result goes beyond Einstein's theory and holds true for more general theories in which the entropy is not proportional to the horizon area.

The need to take into account the surface terms is also dictated by the holographic principle, which postulates a fundamental relationship between the volume and surface degrees of freedom\cite{24s,25s}.

At the phenomenological level, taking into account surface terms in the presence of a horizon leads to a modification of the acceleration equation. The deceleration in expansion caused by volumetric terns(matter) competes with the acceleration generated by surface terms (entropic forces). For a homogeneous and isotropic Universe, the acceleration equation, taking into account the contribution of the surface terms, can be presented in the form
\[\frac{\ddot{a}}{a}=- \frac{4 \pi G}{3} (\rho +3 p)+ \frac{a_{\text{surfaces}}}{d_h}\]
Here $d_h$  is some characteristic distance.

It can be expected that the contribution of the surface term will be of the order
\[\frac{6(2H^2 +H)}{8 \pi} \sim \frac{3}{2 \pi} \left(H^2 + \frac{H}{2} \right)\]
There are various interpretations of the physics in the literature, which leads to the inclusion of surface terms. One of the possible interpretations is the entropic interpretation of the forces generated by the surface terms. Entropy forces naturally lead to slow late-time accelerated expansion of the Universe.

Within this approach
\[a_{surfaces}=a_{entropic}=cH\]
Here  $a_{\text{surfaces}}$ is the additional acceleration generated by the surface terms. There is no dark energy in the proposed approach.  Instead of, the cosmological horizon and its thermodynamic characteristics play a central role. The only assumption of the proposed model is that the horizon has temperature and entropy associated with it. The ideology of the model, of course, is based on the thermodynamics of black holes, where a similar relationship takes place for event horizons with Hawking temperature and Bekenstein entropy.

In entropic cosmology  the apparent horizon of the Universe is usually chosen as the horizon. For a spatially flat Universe, the apparent horizon coincides with the Hubble horizon. Here are some arguments justifying the use of the Hubble horizon as the cosmological horizon. Let's start by answering a fundamentally important question: how close is the Hubble radius $R_H$ to radius of  event horizon  of the observable Universe  $M_H$ (the mass of matter inside the Hubble sphere). Simple calculations
\[R_H =c H^{-1}\]
\[M_H= \frac{4 \pi}{3}R_H^3 \rho\]
\[H^2 = \frac{8 \pi G}{3}\rho \to \rho=\frac{3 H^2}{8 \pi G}\]
\[M_H = \frac{4 \pi}{3}R_H^3 \frac{3 H^2}{8 \pi G}= \frac{c^2 R_H}{2G}\]
\[R_H =\frac{2G M_H}{c^2}=r_g\]
show that the Hubble radius $R_H$ is exactly the same as the gravitational radius $r_g$ of the substance inside the Hubble sphere.

A natural question arises: why, in the presence of the relation
$R_H =\frac{2G M_H}{c^2}=r_g$
Hubble sphere is not an event horizon. The answer is simple: in an expanding Universe, the speed of light is not the limiting speed of receding galaxies.

The speed of the recession of galaxies $V$ obeys Hubble law
$V=HR$ and when crossing the Hubble horizon is equal to the speed of light. the further location of galaxies - outside or inside the Hubble sphere - is determined by the ratio between the speed of galaxies and the speed of the horizon $R_H$,
\[\dot{R}_H =c(1+q)\]	
where $q$ is the deceleration parameter. At $q>0$  (decelerated expansion)), the Hubble sphere behaves like the event horizon of a black hole: the galaxies on the Hubble sphere lag behind it, and the number of galaxies inside the Hubble sphere grows with time. At $q<0$ (accelerated expansion), the Hubble sphere behaves like the event horizon of a white hole: the galaxies on the Hubble sphere are ahead of it, and the number of galaxies inside the Hubble sphere decreases with time - cosmic loneliness. Recall that a white hole is a hypothetical physical object in the Universe, into the region of which nothing can enter.

In entropic cosmology, the cosmological horizon is associated with the Hawking temperature  $T_H$, which we can estimate as
\[T_H = \frac{\hbar H}{2 \pi k_B} \sim 3\times 10^{-30}\quad K\]
The interpretation of this temperature as the Unruh temperature (\ref{ehd4}) makes it possible to estimate the horizon acceleration generated by the entropic force
\[a_{horizon}=\left(\frac{2\pi c k_B T}{\hbar} \right)=cH \approx 3\times 10^{-9}\; \frac{m}{sec^2}\]
We arrive at cosmic acceleration essentially in agreement with observation. This eliminates the need to introduce dark energy.

\section{Entropy cosmology equations}

The introduction of entropic forces (or, equivalently, taking into account the contribution of surface terms when varying the Hilbert-Einstein action) requires a modification of the Friedmann equations. Following\cite{26s,27s,28s,29s,30s,31s,32s}, we consider this procedure for a homogeneous and isotropic Universe with energy exchange between the volume and the Hubble horizon.

From the thermodynamic) point of view, these models can be divided into two classes. The so-called  $\Lambda (t)$-models represent the first class. They are similar to  $\Lambda (t)$-models represent the first class. They are similar to $\Lambda (t)$-generalizations of the SCM, which use a time-dependent cosmological constant, i.e. $\Lambda \to \Lambda (t)$. In the $\Lambda (t)$-models, both the Friedmann equation and the acceleration equation include the same an extra driving term $f_{\Lambda}(t)$. The driving term results in, generally speaking, a non-zero term on the right hand side of the conservation equation, with the exception of the SCM. This non-zero term is considered to be associated with ''reversible'' entropy, for example, due to the exchange of matter (energy) between volume and horizon. In this sense, the Universe is non-dissipative for  $\Lambda (t)$ -models \cite{33s,34s}.

The second class of models is the so-called BV (bulk viscousity) - models, similar to cosmological models that include bulk viscosity\cite{35s,36s,37s,38s,39s,40s}. In BV models the acceleraton equation includes an extra driving term while the Friedman equation does not. This term results in a non-zero term on the right side of the conservation equation, even if the driving term is independent of time. It is assumed that this non-zero term is related to the "irreversible entropy" generated, for example, in the process of matter creation in the gravity field\cite{41s} or in the presence of viscosity. The Universe for BV-models is dissipative. The modified Friedman equations and the conservation equation, suitable for describing both $\Lambda (t)$ and BV-models, can be represented as
\begin{equation}\label{eh10}
  H^2 = \frac{8 \pi G}{3c^2}\rho +f_{\Lambda}(t)
\end{equation}
\begin{equation}\label{eh11}
  \frac{\ddot{a}}{a}=-\frac{4 \pi G}{3}(\rho +3p)+f_{\Lambda}(t)+h_{BH}(t)
\end{equation}
\begin{equation}\label{eh12}
  \dot{\rho}+3H(\rho+p)=\frac{3 \dot{f}_{\Lambda}(t)}{8 \pi G}+ \frac{3 H h_{BV}(t)}{4 \pi G}
\end{equation}
Two extra driving terms  $f_{\Lambda}(t)$  and $h_{BV}(t)$ are introdused phenomenologically\cite{31s}.

The function $f_{\Lambda}(t)$ is used for $\Lambda (t)$-models, and  $h_{BV}(t)$  for BV- models. The first term on the right side of equation (\ref{eh12}) is associated with reversible processes (for example, energy exchange). The second term is generated due to irreversible processes (for example, the  matter creation).

From the system of equations  (\ref{eh10})-(\ref{eh12})one can obtain an equation describing the evolution of the Hubble parameter and, ultimately, evolution of the scale factor $a(t)$
\begin{equation}\label{eh13}
  \dot{H}=-\frac{3}{2}H^2 + \frac{3}{2}f_{\Lambda}(t)+h_{BV}(t)
\end{equation}
Using this equation, we can describe the evolution of the Universe in various cosmological models that differ in the choice of functions   $f_{\Lambda}(t)$ and $h_{BV}(t)$.

An alternative way to introduce the terms of the entropic force (entropic pressure) is to use the effective pressure  $p_{eff}=p+p_e$. The introduction of effective pressure into the acceleration equation and the conservation equation leads to the system
\[H^2 = \frac{8 \pi G}{3 c^2} \rho\]
\[\frac{\ddot{a}}{a}=-\frac{4 \pi G}{3 c^2}(\rho + p_{eff})\]
\begin{equation}\label{eh14}
  \dot{\rho}+3H(\rho +p_{eff})=0
\end{equation}
For the entropic pressure generated by the Bekenstein entropy  $p_e =\frac{c^2}{4 \pi G}H^2$, we find
\[H^2 = \frac{8 \pi G}{3 c^2} \rho\]
\[\frac{\ddot{a}}{a}=-\frac{4 \pi G}{3 c^2}(\rho + p)+H^2\]
\begin{equation}\label{eh15}
  \dot{\rho}+3H(\rho +p)=\frac{3}{\pi}H^3
\end{equation}
Note that in the original model of entropic cosmology\cite{3s}, only the term  $H^2$ or combination $H^2 +H$ was included in both Friedmann equations as extra driving term. This choice corresponded to the assumption of the adiabatic evolution of the Universe or, in other words, represented the model ($f_{\Lambda} (t) \neq 0$, $h_{BV}=0$). The system of equations  (\ref{eh14}) represents the BV-model when  ($f_{\Lambda} (t) = 0$, $h_{BV} \neq 0$).

The conservation equation (\ref{eh14}) with effective pressure can be used to calculate the total entropy, which is the sum of the entropy of the horizon and the entropy of matter inside the Hubble sphere. For such a model, the first law of thermodynamics\cite{42s}
\begin{equation}\label{eh16}
  T dS_m =dE_m + (p+p_e)dV
\end{equation}
Here  $V$ is the volume of the Hubble sphere, $E_m =p V$. From here, we obtain the time derivative of the entropy of matter inside the horizon
\begin{equation}\label{eh17}
  \dot{S}_m=\frac{1}{T}[(\rho +p+p_e)\dot{V}+ \dot{\rho}V ]
\end{equation}
Using the equations
\[\dot{\rho}+ 3H (\rho +p +p_e)=0\]
\begin{equation}\label{eh18}
  H^2 = \frac{8 \pi L^2_{Pl}}{3}\rho
\end{equation}
we find
\[\rho+p+p_e =-\frac{H}{4 \pi L_{Pl}^2}\]
\begin{equation}\label{eh19}
  \dot{\rho}=\frac{3H\dot{H}}{4 \pi L_{Pl}^2}
\end{equation}
Then, substituting  (\ref{eh19}), the Hubble volume $V=\frac{4 \pi}{3 H^3}$ and temperatur $T=\frac{H}{2 \pi}$ in equation  (\ref{eh17}) we obtain the time derivative of the entropy $S_m$ in the form
\begin{equation}\label{eh20}
  \dot{S}_m=\frac{2 \pi \dot{H}}{H^3 L_{Pl}^2}\left(1+\frac{\dot{H}}{H^2}  \right)
\end{equation}
In the literature\cite{43s}, for the entropy of matter inside the horizon the alternative expression is sometimes used
\begin{equation}\label{eh21}
  \dot{S}_m=\frac{2 \pi }{H^2 L_{Pl}^2(2 H^2 +\dot{H})} \left(\frac{\ddot{H}}{2}+2H \dot{H}+ \frac{\dot{H}^3}{H^3} + \frac{2\dot{H}^2}{H} \right)
\end{equation}
If we assume that the first and third terms appear as a consequence of temperature redefinition,
\begin{equation}\label{eh22}
  T=\frac{H}{2 \pi} \to \frac{H}{2 \pi} \left(1+ \frac{\dot{H}}{2H^2}  \right)
\end{equation}
then after removing these terms, formulas  (\ref{eh20}) and (\ref{eh21}) will coincide.

As we saw above, the value  $\frac{H}{2 \pi}$ is usually used as the temperature of the Hubble horizon of a flat FlRW Universe. A free-falling observer at the de Sitter horizon will measure just such a temperature. Our Universe is only a asymptotically de Sitter. The additional term in form (\ref{eh21}) takes into account the necessary corrections.

\section{Emergent Universe}

The discovery of an organic connection between gravity and thermodynamics has stimulated numerous attempts to deprive gravity of its status as a fundamental force. In particular, Verlinde\cite{21s} proposed to consider gravitation as an entropic force. As we saw above, the entropy interpretation of gravity underlies entropic cosmology. Allowing gravity to be treated as an emergent phenomenon, the traditional erenium point assumes that the background space-time is an existing object.

The next radical step was taken by Padmanabhan\cite{44s,45s}, who suggested that space-time itself is an emergent structure.
The main problem with this approach is that it is conceptually difficult to treat time as an emergent structure. However, Padmanabhan managed to overcome this problem in a cosmological context by choosing universal cosmological time as the time variable. In this case, the spatial expansion of the Universe can be described as the emergence of outer space with the development of cosmic time.

As we have already noted, the Unruh effect (Unruh temperature) indicates that space-time can be heated like ordinary matter. From this we can conclude that space-time, like ordinary matter, consists of microscopic degrees of freedom. In models of the Universe with a cosmological horizon, the degrees of freedom can be divided into surface $N_{bulk}$, to the horizon, and volume $N_{sur}$ belonging to the volume inside the Hubble sphere.

According to Padmanabhana's hypothesis, the expansion of the Universe, or, in other words, the emergence of space, is due to a mismatch between the degrees of freedom on the horizon and the volume inside the horizon. As the cosmological time  grows,  $N_{sur}$ increases and approaches to  $N_{bulk}$ (in this case,  $N_{bulk}$ also increases with time), and finally the condition is reached
\begin{equation}\label{eh23}
  N_{sur}=N_{bulk}
\end{equation}
which is called the holographic equipartition. When condition (34) is satisfied, the Universe passes into the de Sitter era. The dynamics of the transition is driven by the equation
\begin{equation}\label{eh24}
  \frac{dV}{dt}=L_{Pl}^2 (N_{sur}-N_{bulk})
\end{equation}
where $V=\frac{4 \pi}{3H^3}$ is the Hubble volume. The number of surface and volume degrees of freedom is defined as
\begin{equation}\label{eh25}
  N_{sur}=\frac{4 \pi}{H^2 L_{Pl}^2},\, N_{bulk}=\frac{|E|}{\frac{1}{2}k_B T}=- \varepsilon \frac{(\rho +3p)}{\frac{1}{2}k_B T}
\end{equation}
Here  $T=\frac{H}{2 \pi}$ is Hawking temperature, is density of  Komar energy\cite{46s,47s} and
\[ \varepsilon =+1 \,\,\text{for} \,\, \rho+p<0 , \quad \varepsilon =-1 \,\, \text{for} \,\, \rho+p>0\]
The viability of the Padmanabhan hypothesis is based on the fact that from (\ref{eh24}) one can derive the Friedmann equations\cite{48s,49s} and dynamic equations for more general theories of gravity\cite{49s}.

Using definitions (\ref{eh25}), equation (\ref{eh24}) can be written as
\begin{equation}\label{eh26}
  -\frac{4 \pi \dot{H}}{H^4}=L_{Pl}^2 \left[\frac{4 \pi}{H^2 L_{Pl}^2}+\frac{16 \pi^2 (\rho+3p)}{3 H^4}  \right]
\end{equation}
The time evolution of the Hubble parameter depends on the Komar energy density  $\rho +3 p$.

\section{Energy flow across the cosmological horizon}

The purpose of this section is to calculate the change in entropy of the Hubble sphere associated with the flow of energy through it. Of course, this problem can be solved only within the framework of a certain axiomatics. In the now classic paper \cite{12s}, T.Jacobson showed that the field equations of GR can be obtained if to require  the validity of the Clausius relation $\delta Q = TdS$ on the local Rindler horizon and the fulfillment of the Bekenstein area-entropy relation on the horizon. Here $\delta Q$ is the energy flux through the local Rindler horizon, and $T$ is the Hawking (Unruh) temperature observed by an accelerated observer near the horizon.

Thus, the problem under consideration contains three fundamental components:
\begin{enumerate}
  \item Clasius relation.
  \item Bekenstein entropy of the horizon.
  \item GR (with the Gilbert-Einstein action.
\end{enumerate}
Of these three assumption,	 only two are independent. T. Jacobson showed that from 1+2 follows 3. The result presented below means that 2 follows from 1+3 (within the form of the modified Friedman equations of entropy cosmology), once again demonstrating the direct relationship between gravity and thermodynamics.

The differential energy flow across the Hubble sphere  -$E$ is equal to the energy change inside the Hubble sphere $dU$ plus the work $dA$, done by expanding the Hubble sphere. This statement can be written in the form of the first law of thermodynamics
\begin{equation}\label{eh27}
  -dE=dU+dA
\end{equation}
Calculate the terms on the right side
\[dU=\rho dV=A_H \rho dr_H\]
\begin{equation}\label{eh28}
  dA=p dV =A_H p dr_H
\end{equation}
Thus
\begin{equation}\label{eh29}
  -dE =A_H (\rho +p)dr_H
\end{equation}
Using Hubble law, we find
\begin{equation}\label{eh30}
  dr_H =vdt=H r_H dt
\end{equation}
And finally
\begin{equation}\label{eh31}
  - dE=A_H (\rho + p)H r_H dt
\end{equation}
A complicated but more rigorous derivation of this formula can be found in\cite{3s,31s,50s,51s}.
To calculate the right side of this equation, we find the multiplier  $(1+\omega)\rho H$ from the modified conservation equation
\begin{equation}\label{eh32}
  \dot{\rho}+3 (1+\omega)\rho H= \frac{3H}{4 \pi G} \left( h_B (t) - \frac{\dot{f}_{\Lambda}(t)}{2 H}\right)
\end{equation}
whence
\begin{equation}\label{eh32b}
  (1+ \omega)\rho H=\frac{1}{3} \dot{\rho}+ \frac{H}{4 \pi G} \left(h_B (t) - \frac{\dot{f}_{\Lambda}(t)}{2 H} \right)
\end{equation}
We then calculate  $- \dot{\rho} /3$  from the modified Friedmann equation
\[H^2 = \frac{8 \pi G}{3} \rho + f_{\Lambda}(t)\]
\begin{equation}\label{eh33}
  -\frac{1}{3} \dot{\rho}=\frac{-2H\dot{H}+\dot{f}_{\Lambda}(t)}{8 \pi G}
\end{equation}
Using the obtained results and  $A_H= 4 \pi r_H^2 =4 \pi \frac{c^2}{H^2}$, we finally find\cite{31s}
\begin{equation}\label{eh34}
  -\dot{E}=\frac{c^5}{G H^2} (\dot{H} +h_B(t))
\end{equation}
For $c=\hbar =1$, $G=L_{Pl}^2$, $h_B =0$
\begin{equation}\label{eh35}
  \dot{E}=\frac{\dot{H}}{H^2L_{Pl}^2}
\end{equation}
The resulting expression  (\ref{eh35}) for $ \dot{E}$, allows us to calculate the time derivative of entropy using the definition
\begin{equation}\label{eh36}
  T dS_h =-dE
\end{equation}
The minus sign is due to the fact that we considered the energy leaving the volume. For $T=H/2 \pi$ we find
\begin{equation}\label{eh37}
  \dot{S}_h=\frac{2 \pi \dot{H}}{H^3 L_{Pl}^2}
\end{equation}
which exactly coincides with the time derivative of the Bekenstein entropy.

Due to the importance of this result, we calculate the energy flux across Hubble horizon in the framework of emergent cosmology\cite{50s,52s,53s}.

The energy at the Hubble horizon can be given as
\begin{equation}\label{eh38}
  \varepsilon = N \frac{1}{2}k_B T
\end{equation}
where  $N$ is the number of degrees of freedom on the Hubble sphere which is written as
\begin{equation}\label{eh39}
  N=\frac{4 S}{k_{B}}
\end{equation}
Using the expression $T=\frac{\hbar H}{2 \pi k_B}$ for the temperature, we find
\begin{equation}\label{eh40}
  \varepsilon =\frac{4S_{BH}}{k_B} \cdot \frac{1}{2}k_B T=2 S_{BH}T
\end{equation}
The equation $\varepsilon =2 S_{BH}T$  was explored by Padmanabhan\cite{54s}. Substituting the expressions for entropy and temperature, we obtain
\begin{equation}\label{eh41}
  \varepsilon = 2 \frac{\pi k_B c^5}{\hbar G}\frac{1}{H}\cdot \frac{\hbar H}{2 \pi k_B}= \frac{c^5}{GH}
\end{equation}
Whence
\begin{equation}\label{eh42}
  \dot{\varepsilon}= \frac{c^5}{G} \left( -\frac{\dot{H}}{H^2}\right)=\frac{\dot{H}}{H^2 L^2_{Pl}}
\end{equation}
Comparing the expression  (\ref{eh34}) with the expression  (\ref{eh28}) obtained above for   $-\dot{E}$ in $\Lambda (t)$-model, we find
\begin{equation}\label{eh43}
  -\dot{E}=\dot{\varepsilon}=\frac{c^5}{G} \left(- \frac{\dot{H}}{H^2}\right)=-\frac{\dot{H}}{H^2 L_{Pl}^2}
\end{equation}
This coincidence increases confidence in the key result, the entropy of the Hubble horizon, since it was achieved by fundamentally different methods. A more detailed interpretation of this relation can be found in\cite{50s,55s}.

Note that when deriving relation (\ref{eh35}) ad hoc, the Hawking temperature $T=\frac{H}{2 \pi k_B}$ was used as the temperature of the cosmological horizon. Is a scheme based on such an assumption self-consistent within the considered axiomatics? To answer this question, we again use the Clausius relation, but this time we fix the entropy (the Bekenstein entropy) rather than the temperature. In other words, we will consider a simplified version of Jacobson's formulation\cite{12s} with the goal of obtaining the horizon temperature.

Using the Clausis relation and the expression for the energy flux through the Hubble horizon $dE=A(\rho + p)dt$, the time derivative of the horizon entropy can be represented as
\begin{equation}\label{ehn44}
  \dot{S}_h =\frac{A (\rho + p)}{T_h}
\end{equation}
From here we find
\begin{equation}\label{ehn45}
  \dot{A}=4G\frac{(\rho +p)}{T}A
\end{equation}
Considering that $\dot{H}=-4G \pi (\rho +p)$ for the horizon temperature we obtain
\begin{equation}\label{ehn46}
  T_h = - \frac{\dot{H}}{\pi}\frac{A}{\dot{A}}=\frac{H}{2 \pi}
\end{equation}
We emphasize that the above arguments are not a conclusion of the Hawking temperature, but serve as a test of the self-consistency of the considered axiomatics. The latter dictates a rigid relationship between the entropy and the temperature of the horizon: the choice of one of these thermodynamic characteristics uniquely determines the second.

\section{Generalized entropy of the cosmological horizon}

Until now, it is not entirely clear why the entropy of a black hole is proportional to the area of the event horizon, and not to its volume. In classical thermodynamics, the entropy of a system, being an extensive quantity, is proportional to its mass (volume). Obviously, this question is even more concerned with cosmological horizons (and entropy cosmology in general), since in this case the number of alternative possibilities increases dramatically. Let's start with the simplest generalization of the Bekenstein entropy

\noindent
\begin{equation} \label{ehn59}
S_{r2} =\frac{k_{B} c^{3} }{\hbar G} \frac{A_{H} }{4} =\frac{k_{B} c^{3} }{\hbar G} \frac{4\pi r_{H}^{2} }{4} =\frac{\pi k_{B} c^{3} }{\hbar G} r_{H}^{2}
\end{equation}
This entropy is proportional to $r_{H}^{2} $ .  Natural generalizations of this form as entropy of the cosmological horizon   are the so-called volumetric entropy $S_{r3} {\rm \sim }r_{H}^{3} $   and entropy of the fourth degree\cite{29s,30s}. The entropy $S_{r3} $ is a generalization of the black hole entropy to the case of non-additive Tsallis statistics\cite{56s,57s}. Entropy $S$ can be interpreted as the entropy associated with additional dimensions.

The general form of the reversible part of the entropy on the Hubble horizon can be defined as\cite{23s}
\begin{equation} \label{ehn60}
S_{m} =\frac{\pi k_{B} c^{3} }{\hbar G} L_{m} r_{H}^{m}
\end{equation}
where $m=2,3,4$ in the cases of Bekenstein, Tsallis and quartic entropy. $L_{m} $ is the free parameter. $L_{m} =1$ for , for , for $m=2$, $L_{m} =\xi $ for $m=3$, $L_{m} =\chi $ for  $m=4$. Here $\xi $ and $\chi $ are non-negative parameters. Entropy force generated by entropy change
\begin{equation} \label{ehn61}
F_{m} =-T\frac{dS}{dr_{H} } =-\frac{c^{4} }{G} \frac{mL_{m} }{2} r_{H}^{m-2}
\end{equation}
Hence, the expression for the entropy pressure generated by the entropic force on the cosmological horizon can be written as
\begin{equation} \label{ehn62}
p_{e} =\frac{F_{m} }{A} =\left(\frac{c^{m} mL_{m} }{8\pi G} \right)H^{4-m}
\end{equation}
The Friedmann equations for the model with entropic forces  \eqref{ehn60}  can be written in the form
\begin{equation} \label{ehn63}
H^{2} =\frac{8\pi G}{3} \rho +\left(\frac{c^{m-2} mL_{m} }{2} \right)H^{4-m}
\end{equation}
\begin{equation} \label{ehn64}
\frac{\ddot{a}}{a} =\frac{4\pi G}{3} (\rho +3p)+\left(\frac{c^{m-2} mL_{m} }{2} \right)H^{4-m}
\end{equation}
Using equations \eqref{ehn63} and \eqref{ehn64} we obtain the conservation equation
\begin{equation} \label{ehn65}
\dot{\rho }+3H(\rho +p_{eff} )=0
\end{equation}
where $p_{eff} =p+p_{e} $  . Or in the equivalent form
\begin{equation} \label{ehn66}
\dot{\rho }+3H(\rho +p)=\left(\frac{3c^{m-2} mL_{m} }{2} \right)\left(\frac{4-m}{2} \right)H^{3-m} \dot{H}
\end{equation}
The entropy in the form $S_{r2} ,S_{r3} ,S_{r4} $ leads to the appearance in the Friedmann equation of extra driving term $H^{2} ,H$ and a constant, respectively. Let us note that the structure of equations \eqref{ehn63} and \eqref{ehn64} corresponds to the $\Lambda (t)$-model: the  extra terms in the Friedmann equation \eqref{ehn63} and the acceleration equation \eqref{ehn64} coincide. This is due to the use of "reversible" entropy, which corresponds to the exchange of energy between the horizon and the volume of the Universe. As physics progresses, it includes into consideration systems that differ sharply both in characteristic scales and in structural features. Working with such a variety involuntarily leads to the conclusion that there is no single implementation of the definition of entropy, since it depends on the physical system under consideration. This situation has led to the emergence of a number of definitions of entropy. Each of which is adapted to solve specific physical problems. Here are some definitions of entropy that are potentially useful for use in entropic cosmology.

The Tsallis entropy\cite{56s} proposed for the case of non-extensive (non-additive) systems is a generalization of the standard Boltzmann-Gibbs entropy
\begin{equation} \label{ehn67}
S_{T} =\frac{A_{0} }{4G} \left(\frac{A}{A_{0} } \right)^{\delta }
\end{equation}
where $A_{0}$ and $\delta$ are constants. For $\delta =1$ the Tsallis entropy reduces to the Bekenstein entropy

\noindent The R\'{e}nyi entropy\cite{58s}, originally introduced in information theory as a generalization of the Shannon entropy, turned out to be useful for describing systems with long-range forces (in particular, in gravity)
\begin{equation}\label{ehn68}
  S_{R} =\frac{1}{\alpha } (1+\alpha S)
\end{equation}

Here (and in further definitions of entropies) $S$ is the Bekenstein entropy, $\alpha $is parameter. At $\alpha \to 0$ , the R\'{e}nyi entropy coincides with the Bekenstein entropy.

The Barrow entropy is given by the formula\cite{59s}
\begin{equation} \label{ehn69}
S_{B} =\left(\frac{A}{4G} \right)^{1+\frac{\Delta }{2} }
\end{equation}
Inspired by illustrations of the Covid-19 virus, Barrow argued that quantum gravitational effects could transform the surface of the event horizon into a fractal structure described by a parameter $\Delta $ . Two characteristic values of this parameter: $\Delta =0$ corresponds to the Bekenstein entropy, and $\Delta =1$ leads to an increase in the fractal dimension of the horizon surface by one.

Information about other forms of entropy used in cosmology can be found in\cite{60s,60ss},

\noindent The variety of forms of entropy used in cosmology raises a natural question: is there a generalized form of entropy that generalizes well-known entropy functions such as Tsallis, Renyi, Barrow, and others? If so, what is the minimum number of parameters that define such a construct? A positive answer to this question was given in\cite{60s}. The authors proposed a 4-parameter construction that can generalize all the above (and a number of unmentioned) entropies. This generalized entropy  is given by
\begin{equation} \label{ehn70}
S_{g} (\alpha _{+} ,\alpha _{-} ,\beta ,\gamma )=\frac{1}{\gamma } \left[\left(1+\frac{\alpha _{+} }{\beta } S\right)^{\beta } -\left(1+\frac{\alpha _{-} }{\beta } S\right)^{-\beta } \right]
\end{equation}
where $\alpha _{+} ,\alpha _{-} ,\beta ,\gamma$ are parameters that are assumed to be positive. Generalized entropy is reduced to the forms of entropy mentioned above with an appropriate choice of parameters. So, for example, for $\alpha _{+} \to \infty$  and $\alpha _{-} =0$  the generalized entropy reduces to $S_{g} =\frac{1}{\gamma } \left(\frac{\alpha _{+} }{\beta } S  \right)^{\beta } $.  If we choose $\gamma =\left(\frac{\alpha _{+} }{\beta } \right)^{\beta }$  get $ S_{g} =S^{\beta }$. For $\beta =\delta$  or $\beta =1+\frac{\Delta }{2}$  we reproduce the Tsalis entropy or Barrow entropy, respectively. The use of generalized entropy (\ref{ehn70}) leads to a number of interesting consequences for cosmological evolution\cite{60s}.

\section{Entropic cosmology and laws of thermodynamics}

The expansion of the Universe is traditionally described in terms of GR or its numerous generalizations. An analysis of the thermodynamics of this process opens up alternative possibilities for understanding the evolution of the Universe. Both approaches describe the same reality and therefore there must be a deep relationship between them. In this section, we will directly trace the connection between the laws of thermodynamics and cosmological models built on the basis of GR.

\subsection{First law of thermodynamics}

Let's start by analyzing the relationship between the first law of thermodynamics and Friedmann equations. In the previous section, it was shown that fixing the Clasius relation and the Friedmann equations uniquely determine the entropy of the cosmological horizon. Here we want to solve the inverse problem: using the first law of thermodynamics and fixing the entropy of the horizon, we obtain the Friedmann equations for the case of arbitrary spatial curvature\cite{61s,62s}.

Recall that the main characteristics of a Schwarzschild black hole such as energy $\left(E=M\right)$, entropy and temperature are related by a thermodynamic identity, usually called the first law of black hole dynamics $TdS=dE$. The Schwarzschild metric is a vacuum solution without pressure. This explains the absence of a term $PdV$. The metric has only one parameter $M$, so changes in all physical characteristics can be associated with a change in mass. If we want to use the relation $TdS=dE$ (treating it as the first law of thermodynamics) in a more general cosmological context, we should take into account the work of changing the cosmological horizon along with the energy change associated with crossing the horizon by matter (galaxies). Following\cite{62s} let us calculate the energy flux $dE$through the horizon of the n$\left(n+1\right)$ -dimensional FLRW Universe, the metric of which can be represented in the form \eqref{ehd1}, and the radius of the visible horizon is determined by relation \eqref{ehd2}.

\noindent In order to take into account the work associated with  expansion of the Universe, we define the work density\cite{63s,64s,65s}
\begin{equation} \label{ehn71}
W=-\frac{1}{2} T^{ab} h_{ab}
\end{equation}
and the energy-momentum vector
\begin{equation} \label{ehn72}
\Psi _{a} =T_{a}^{b} \partial _{b} \tilde{r}+W\partial _{a} \tilde{r}
\end{equation}
Where $T_{ab} $ is the projection of the $(n+1)$-dimensional energy-momentum tensor of an ideal fluid filling the FLRW Universe in the normal direction of the $(n-1)$-dimensional sphere. The work density on the apparent horizon should be considered as the work done to change the apparent horizon, and the energy flow vector as the total energy crossing the horizon.

\noindent Let's proceed to the calculation of the heat flux $\delta Q$ across  the apparent horizon during an infinitely small time interval $dt$. This heat flux represents the change in energy within the apparent horizon, i.e $\delta Q=-dE$. The energy-momentum tensor  $T_{\mu \nu } $in the Universe filled with an ideal fluid has the form $T_{\mu \nu } =\left(\rho +p\right)u_{\mu } u_{\nu } +pg_{\mu \nu } $. Using \eqref{ehn71}, \eqref{ehn72} we find the components of the energy-momentum vector

\noindent
\begin{equation} \label{ehn73}
\Psi _{a} =\left(-\frac{1}{2} \left(\rho +p\right)H\tilde{r},\frac{1}{2} \left(\rho +p\right)a\right)
\end{equation}
The amount of energy crossing the horizon during time $dt$
\begin{equation} \label{ehn74}
-dE\equiv A\Psi =A\left(\rho +p\right)H\tilde{r}_{A} dt
\end{equation}
This formula coincides with relation \eqref{eh29} obtained in a less rigorous way.

Let us assume that the visible horizon is characterized by the Bekenstein entropy $S=\frac{A}{4G} $ and the Hawking temperature $T=\frac{1}{2\pi \tilde{r}_{A} } $. Substituting these quantities into the first law of thermodynamics in the form $-dE=TdS$ we find
\begin{equation} \label{ehn75}
\dot{H}-\frac{k}{a^{2} } =\frac{8\pi G}{n-1} \left(\rho +p\right)
\end{equation}
Using the conservation equation$\dot{\rho }+nH\left(\rho +p\right)=0$, we finally obtain
\begin{equation} \label{ehn76}
H^{2} +\frac{k}{a^{2} } =\frac{16\pi G}{n(n-1)} \rho ,
\end{equation}
Equation \eqref{ehn76} is the Friedmann equation describing the $(n+1)$-dimensional FLRW Universe with spatial curvature. A similar procedure can be used to demonstrate the connection between the first law of thermodynamics and the law of  FLRW Universe  in emergent cosmology\cite{66s,67s}.

\subsection{The second law of thermodynamics}

Let us now turn to the consideration of the question of the fulfillment of the second law of thermodynamics in entropic cosmology. Note that a negative answer to this question will deprive entropic cosmology of any prospects. Recall that initially entropic cosmology arose as an attempt to transfer the laws of thermodynamics of black holes to the thermodynamics of the Universe. The success or failure of this attempt largely depends on whether the second law of thermodynamics holds on cosmological scales.

Estimates show that the entropy of the Universe (we mean that part of the Universe that is in causal contact with us) is dominated by the entropy of the cosmological horizon. Supermassive black holes and cosmic microwave radiation lag behind by 18 and 33 orders of magnitude, respectively. All other sources of entropy make a much smaller contribution\cite{20s}.

Therefore, the analysis of the behavior of the horizon entropy during the evolution of the Universe represents the greatest interest. According to the currently dominant point of view, our Universe, described by GR, behaves like an ordinary thermodynamic body. This, in particular, means that in the process of the Hubble expansion, the Universe evolves to the equilibrium state of maximum entropy, subject to the restrictions
\begin{equation} \label{ehn77}
\dot{S}\ge 0,always,
\end{equation}
\begin{equation} \label{ehn78}
\ddot{S}<0,at\; least\; lomg\quad run
\end{equation}
Here  $S$ represents the total entropy of the universe, which can be approximated by the entropy of the horizon and dots denote derivatives with respect to cosmological time.

We can consider any closed surface (such as the event horizon or apparent horizon) as the thermodynamic boundary of the system, through which energy and matter can enter or leave the system. Here we consider the apparent horizon with radius \eqref{ehd3} as such a surface. Bekenstein entropy in $n$-dimensional case
\begin{equation} \label{ehn79}
S=\frac{A_{n+1} }{4L_{P}^{n-1} }
\end{equation}
where $A_{n+1} =n_{\Omega n} \tilde{r}_{A} ^{n-1} $ for $n\ge 3$ is the area of the apparent horizon, $\Omega _{n} $is the volume of a unit $n$-dimensional sphere. Let us check whether the entropy \eqref{ehn79} satisfies the constraints \eqref{ehn77} and \eqref{ehn78} \cite{68s}.

For this purpose, we need its first and second time derivatives. Using \eqref{ehn77}, we find
\begin{equation} \label{ehn80}
\dot{S}=\frac{n(n-1)\Omega _{n} }{4L_{P}^{n-1} } \tilde{r}_{A} ^{n-2} \dot{\tilde{r}}_{A}
\end{equation}
\begin{equation} \label{ehn81}
\ddot{S}=\frac{n(n-1)\Omega _{n} }{4L_{P}^{n-1} } \tilde{r}_{A} ^{n-3} \left[\left(n-2\right)\dot{\tilde{r}}_{A} ^{2} +\tilde{r}_{A} \ddot{\tilde{r}}_{A} \right]
\end{equation}
From \eqref{ehn78} it follows that the requirement $\dot{S}\ge 0$ is equivalent to the condition $\dot{\tilde{r}}_{A} \ge 0$ . Using definition \eqref{ehd2}, the Friedman equation and the conservation equation, we obtain
\begin{equation} \label{ehn82}
\dot{\tilde{r}}_{A} =\frac{n}{2} Hr_{A} (1+\omega )
\end{equation}
where $\omega $is the equation of state parameter $p=\omega \rho $. For $\omega \ge -1$ the derivative $\dot{S}$will be non-negative, which is a necessary condition for the fulfillment of the second law of thermodynamics.

Now let us check whether the entropy \eqref{ehn77} reaches its maximum value during the transition to thermodynamic equilibrium, i.e., whether the condition $\ddot{S}<0$ is satisfied in the asymptotic limit. Entropy maximization also requires the fulfillment of the inequality
\begin{equation} \label{ehn83}
\left|\left(n-2\right)\dot{\tilde{r}}_{A} ^{2} \right|<\left|\tilde{r}_{A} \ddot{\tilde{r}}_{A} \right|
\end{equation}
at least in the last stage of evolution. Differentiating \eqref{ehn82}, we find
\begin{equation} \label{ehn84}
\ddot{\tilde{r}}_{A} =\frac{n}{2} r_{A} \left[\frac{n}{2} (1+\omega )^{2} H^{2} +(1+\omega )\dot{H}+\dot{\omega }H\right]
\end{equation}
In the final de Sitter stage of evolution $\omega \to -1$ and $\dot{\omega }<0$ , therefore $\ddot{\tilde{r}}_{A} <0$. On the other hand, according to \eqref{ehn82} in this limit $\dot{\tilde{r}}_{A} \to 0$. Therefore, the inequality $\ddot{S}<0$ holds for $t\to \infty $, ensuring entropy maximization. Thus, in GRthe entropy of the cosmological horizon of the Universe with arbitrary spatial curvature will never increase indefinitely. Above, we considered the problem of fulfilling the second law of thermodynamics in entropy cosmology, taking into account only the entropy of the cosmological horizon. In some problems, for example, in models with the birth of matter, it is necessary to include in the consideration the entropy of matter inside the cosmological horizon. In the next section, we will consider such a situation.

We emphasize that the above evidence for the fulfillment of the second law of thermodynamics in the framework of entropic cosmology is purely model. This is due to the fact that the latter is only a cosmological model, the adequacy of which should be based on observations.  These observations should address at least two major pain points of entropic cosmology.  Firstly, entropic cosmology implicitly uses the assumption that the Universe is an object subject to the laws of classical thermodynamics. Unlike black holes, this statement for the Universe is just a hypothesis that needs to be confirmed. Secondly, the asymptotic $(t\to \infty )$ value of the entropy of the Universe causes concern. These fears are related to the fact that the entropy of systems with the dominance of Newtonian gravity can grow indefinitely\cite{69s,70s}, excluding entropy maximization due to violation of the condition \eqref{ehn78}. In GR the unlimited growth of entropy is prevented by the formation of black holes. In the process of evolution black hole comes into equilibrium with its own radiation, stabilizing the entropy\cite{71s}.

However, the transfer of the results obtained for black holes to the Universe requires caution. Can an analysis of the Hubble expansion eliminate these concerns? A positive response to this question was received in\cite{72s}. The analysis performed in this work was based on the following assumptions:
\begin{enumerate}
  \item The Universe on sufficiently large scales is well described by FLRW metric
  \item The entropy of the Universe is dominated by the entropy of the cosmological horizon
  \item The entropy of the horizon is proportional to its area
\end{enumerate}
These basic assumptions have been considered in conjunction with well-established observational data.  As a result, the authors came to the conclusion that the entropy of the Universe, like the entropy of an ordinary thermodynamic system, tends to maximization.

\subsection{Third law of thermodynamics}

The third law of thermodynamics states that the entropy of a system should approach a constant value (C) at absolute zero temperature
\begin{equation} \label{ehn85}
S\left(T \to 0\right) \to C
\end{equation}
As we saw above, the Clausius relation and the Friedmann equations dictate the Bekenstein entropy to the cosmological horizon. Does this entropy satisfy requirement \eqref{ehn85}\cite{73s}?

For a Schwarzschild black hole with mass$M$, the Hawking temperature is

\begin{equation} \label{ehn86}
T_{H} =\frac{\partial E}{\partial S} =\frac{\partial M}{\partial S} =\frac{1}{8\pi M}
\end{equation}
($c=\hbar =G=k_{B} =1$) It is obvious that such a dependence $T_{H} \left(M\right)$ leads to violation of the third law of thermodynamics. A similar problem occurs for the black holes with Kerr-Newman and Reissner-Nordstr metric\cite{74s,75s,76s,77s}.

\noindent It seems natural to study the status of the third law of thermodynamics in the context of generalized entropies, which have successfully proven themselves in solving a number of cosmological problems. Consider this possibility using the Tsallis entropy \cite{56s} in form
\begin{equation} \label{ehn87}
S_{T} =\gamma A^{\delta }
\end{equation}
where  $\gamma $and  $\delta $ are two free parameters. The horizon temperature corresponding to this entropy in classical thermodynamics
\begin{equation} \label{ehn88}
T=\frac{\partial M}{\partial S} =\frac{1}{2\delta \gamma \left(16\pi \right)^{\delta } M^{2\delta -1} }
\end{equation}
The third law of thermodynamics is satisfied whenever  $0<\delta <1/2$, since in this case

\noindent
\begin{equation} \label{ehn89}
S\to 0\quad \left(M\to 0,T\to 0\right)
\end{equation}
Note that if we use the Hawking temperature $\left(T_{H} =\frac{1}{8\pi M} \right)$instead of the temperature given by  $T=\frac{\partial M}{\partial S} $, then the Tsallis entropy in this case is $S\propto T^{-2\delta } $ . So the third law is satisfied only for $\delta <0$   and for this case $S_{T} ,T\to 0$ if  $M\to \infty $. Thus we arrive at a contradiction between two definitions of temperature. The reason for this contradiction\cite{78s} will be discussed in the next section.

To conclude this section, let us dwell on one intriguing feature of the evolution of black holes and, as a consequence, cosmological horizons\cite{79s}. One alternative formulation of the third law of thermodynamics states that for any object, including black holes, reaching zero temperature requires an infinite number of steps associated with erasing information about the formation of a black hole.The evolution of a black hole is dominated by two competing processes. First, the growth of entropy due to the absorption of the surrounding matter. This process leads to a decrease of the temperature of the black hole. Secondly, due to the Hawking radiation, the black hole reduces its mass (evaporates) and, therefore, will increase the temperature. For entropy, in this case, the generalized second law of thermodynamics is satisfied, according to which the total entropy (of a black hole and Hawking radiation) does not decrease. The evaporation process dominates for an isolated black hole. A natural question arises, what happens in the case of dominance in the case of dominance of the process of absorption of matter\cite{79s}. Do black holes keep growing indefinitely? Is there an upper thermodynamic limit on the size of a black hole?

The only way to lower the temperature of a black hole is to increase its mass by accreting the surrounding matter. However, it is impossible to eliminate the quantum mechanical fluctuations of matter falling into a black hole. These fluctuations cause black holes to have a finite lower temperature and, as a consequence, a finite event horizon radius. The estimate of the upper limit of the radius of the cosmological horizon of the Universe is close to the size of the Hubble  horizon $10^{26} $ m.

\section{Thermodynamically consistent entropic cosmology}

The temperature of the horizon of the Universe (for a given entropy) can be obtained from a clear physical requirement --- the preservation of the Legendre structure of thermodynamics \cite{80s,81s}. Cosmological models that do not satisfy this requirement should be considered unsuitable for describing cosmological evolution. A natural question arises: does the simultaneous use of the Bekenstein entropy and the Hawking temperature violate the Legendre structure of thermodynamics?

Consider the free energy of an arbitrary $d$-dimensional system
\begin{equation} \label{ehn90}
G(V, T, p, \mu , \ldots)=U(V, T, p, \mu , \ldots)-TS(V, T, p, \mu , \ldots)+pV-\mu N(V, T, p, \mu , \ldots)-\ldots
\end{equation}
where $T,p,\mu$ - temperature, pressure, chemical potential and  $U,S,V,N$ are internal energy, entropy, volume and number of particles, respectively. Three different types of variables can be distinguished in this Legendre transformation (see \cite{80s} and references therein) Extensive variables $S,V,N, ...$, which are scaled as $V=L^d$, where $L$ is the characteristic linear size of the $d$ -dimensional system, Variables characterizing the external conditions in which the system is placed   $T,p,...$, scaling as  $L^{\theta}$. The energy variables $G,U$  are scaled as  $L^{\varepsilon}$.
From (\ref{ehn90}) we find
\begin{equation}\label{ehn91}
  \varepsilon = \theta + d
\end{equation}
where standard thermodynamics (with short-range interaction) corresponds to $ \theta = 0$.
Consider now a Schwarzschild $(3+1)$-dimensional black hole. In this case, energy scales as mass, which in turn scales as $L$. Therefore $\varepsilon  = 1$, and
\begin{equation}\label{ehn92}
  \theta  = 1 - d
\end{equation}
If we physically identify a black hole with its event horizon surface, then it should be considered as a system with $d = 2$ , then $\theta =- 1$ . This restores the usual Bekenstein-Hawking scaling $T \sim 1/L \sim 1/M$.  However, if the black hole is to be considered as a system with $d = 3$, we have to use ${S_{\delta  = 3/2}}$ entropy. Hence $\theta  = - 2$, i.e.  $T$ is scaled as $T \sim 1/{L^2} \sim 1/{M^2}$. This is key moment:  unless we want to break the Legendre structure of thermodynamics (Eq. (\ref{ehn78})), the Hawking temperature cannot be the temperature of  the cosmological model (unless the chosen entropy is scaled as an area).

Simultaneous use for a black hole of the Hawking temperature and any entropy different from the Bekenstein entropy leads to violation of Legendre's thermodynamic structure. When dealing with the entropy force of a cosmological model based on entropies other than the Bekenstein-Hawking model, there are two possibilities: (i) to keep the Hawking temperature for the horizon, which is contrary to thermodynamics, or (ii) to work with a model consistent with thermodynamics by changing the temperature. Considering thermodynamics as one of the most fundamental physical theories, the second option seems preferable.

\section{Model implementation of entropy cosmology}

In this section, two model implementations of entropy cosmology are considered as examples and an assessment is made of their competitiveness in comparison with traditional cosmological models.

\subsection{Horizon thermodynamics in the  entropic cosmology model}

In this subsection, we consider some model implementations of entropic cosmology as examples and evaluate how successfully they can compete with traditional cosmological models.

To explain the accelerated expansion of the Universe, any cosmological model within the framework of  GR  must introduce additional driving terms into the Friedmann equation. In SCM, this function is performed by the cosmological constant. As we saw above (see section 5), the entropic cosmology scenario relates the appearance of additional driving terms with entropic forces. As a rule, these driving terms are functions of the Hubble parameter (proportional  ${H^2}$ for the Bekenstein entropy and $H$ for the Tsallis entropy. Using the equipartition law in \cite{82s}, a more general case was considered -- the driving term is proportional to ${H^\alpha }$ ( $\alpha $ is the real number). As we saw above, the models entropic cosmology is conveniently divided into two classes $\Lambda (t)$ and $B$. We start by considering $\Lambda (t)$-model, for which the Friedmann equation, the accelerator equation and the conservation equation, and the evolution equation for the Hubble parameter are written as
\begin{equation}\label{ehn93}
  {H^2} = \frac{{8\pi G}}{3}\rho  + {f_\Lambda }(t)
\end{equation}
\begin{equation}\label{ehn94}
  \frac{{\ddot a}}{a} =  - \frac{{4\pi G}}{3}(\rho  + 3p) + {f_\Lambda }(t)
\end{equation}
\begin{equation}\label{ehn95}
  \dot \rho  + 3H(\rho  + p) =  - \frac{{3{f_\Lambda }(t)3}}{{8\pi G}}
\end{equation}
\begin{equation}\label{ehn96}
  \dot H =  - \frac{3}{2}(1 + w){H^2} + \frac{3}{2}(1 + w){f_\Lambda }(t)
\end{equation}
We choose the driving term in the form [31]\cite{31s}
\begin{equation}\label{ehn97}
  {f_\Lambda }(t) = {f_\alpha }(H) = {C_\alpha }H_0^2{\left( {\frac{H}{{{H_0}}}} \right)^\alpha }
\end{equation}
Here $\alpha $ and  ${C_\alpha }$ are independent dimensionless free parameters independent of time. It is assumed that $0 \le {C_\alpha } \le 1$. Model (\ref{ehn97}) reproduces the driving term in the form of a constant, $H$ and  ${H^2}$ for  $\alpha  = 0,1,2$ respectively. Below, for simplicity, we will consider the Universe filled with non-relativistic matter with $w = 0$. In this case, the evolution equation for the Hubble parameter takes the form
\begin{equation}\label{ehn98}
  \dot H = \frac{3}{2}{H^2}\left( {1 - {C_\alpha }{{\left( {\frac{H}{{{H_0}}}} \right)}^\alpha }} \right)
\end{equation}
Solutions to this equation:
for $\alpha  \ne 2$
\begin{equation}\label{ehn99}
  {\left( {\frac{H}{{{H_0}}}} \right)^{2 - \alpha }} = (1 - {C_\alpha }){\left( {\frac{a}{{{a_0}}}} \right)^{ - \frac{3}{2}(2 - \alpha )}} + {C_\alpha }
\end{equation}
for $\alpha  = 2$
\begin{equation}\label{ehn100}
  \frac{H}{{{H_0}}} = {\left( {\frac{a}{{{a_0}}}} \right)^{\frac{3}{2}(1 - {C_\alpha })}}
\end{equation}
Let us dwell on the physical meaning of the parameter ${C_\alpha }$. Passing to the SCM limit in solution (\ref{ehn99}), we obtain
\begin{equation}\label{ehn101}
  {\left( {\frac{H}{{{H_0}}}} \right)^2} = (1 - {C_\alpha }){\left( {\frac{a}{{{a_0}}}} \right)^{ - 3}} + {C_\alpha }
\end{equation}
Comparing this expression with the Friedman equation in SCM
\begin{equation}\label{ehn102}
  {\left( {\frac{H}{{{H_0}}}} \right)^2} = (1 - {\Omega _\Lambda }){\left( {\frac{a}{{{a_0}}}} \right)^{ - 3}} + {\Omega _\Lambda }
\end{equation}
we see that the parameter ${C_\alpha }$ in entropic cosmology can be given the meaning of the equivalent of the relative density of dark energy  ${\Omega _\Lambda }$ in the SCM.  The meaning of the restriction $0 \le {C_\alpha } \le 1$ made above becomes obvious
Let us now discuss the problem of finding the model parameters $\alpha $ and ${C_\alpha }$.  There  are two approaches to determining the parameters of cosmological models. The first, extremely time-consuming approach is to sequentially study the parametric space in order to find the optimal set of parameters. The absence of unambiguous criteria for the concept ''optimal set'' (especially in the case of multi-parameter models) and the need to use significant computing resources reduce the effectiveness of this approach.  Nevertheless, ''blind'' enumeration of parameters remains the dominant method for finding the parameters of cosmological models.

Let us now formulate an alternative approach to finding the model parameters \cite{83s}. The essence of the proposed approach is extremely simple. Let the object under study be described by some evolution equation with n free parameters. Assuming that the dynamic variables are multiply differentiable functions of time, we differentiate the original evolution equation n times. The resulting system can be used to express free parameters in terms of time derivatives of dynamic variables. The latter can be considered as a set of kinematic (cosmographic) parameters available for observation. In the case of interest to us, such a set consists of time derivatives of the scale factor $a(t)$
\[H(t) = \frac{1}{a}\frac{{da}}{{dt}},\]
\[q(t) = \frac{1}{a}\frac{{{d^2}a}}{{d{t^2}}}{\left[ {\frac{1}{a}\frac{{da}}{{dt}}} \right]^{ - 2}},\]
\begin{equation}\label{ehn103}
j(t) = \frac{1}{a}\frac{{{d^3}a}}{{d{t^3}}}{\left[ {\frac{1}{a}\frac{{da}}{{dt}}} \right]^{ - 3}},
\end{equation}
\[s(t) = \frac{1}{a}\frac{{{d^4}a}}{{d{t^4}}}{\left[ {\frac{1}{a}\frac{{da}}{{dt}}} \right]^{ - 4}},\]
\[l(t) = \frac{1}{a}\frac{{{d^5}a}}{{d{t^5}}}{\left[ {\frac{1}{a}\frac{{da}}{{dt}}} \right]^{ - 5}}\]	
An attractive aspect of this approach to determining parameters is the need to solve a system of algebraic rather than differential equations. It should be emphasized that the relations between model parameters and kinematic parameters obtained in this way are exact.

Applying the above scheme to a two-parameter model $\{ \alpha ,C\} $ (the index $\alpha $ of the parameter ${C_\alpha }$ can be omitted), we consider the system of equations
\[\dot H + \frac{3}{2}{H^2} = A{H^\alpha },\]
\begin{equation}\label{ehn104}
  H\ddot H + 3{H^2}\dot H = \alpha A{H^\alpha }\dot H,
\end{equation}
\[A \equiv \frac{3}{2}CH_0^{2 - \alpha }\]
Using
\[\dot H =  - {H^2}(1 + q),\]
\begin{equation}\label{ehn105}
  \ddot H = {H^3}(j + 3q + 2)
\end{equation}
we find a solution to the system (\ref{ehn104})
\[\alpha  = \frac{{\frac{{\ddot H}}{{H\dot H}} + 3}}{{\frac{{\dot H}}{{{H^2}}} + \frac{3}{2}}} = \frac{{2(1 - {j_0})}}{{(1 + {q_0})(1 - 2{q_0})}},\]
\begin{equation}\label{ehn106}
  C = \frac{2}{3}\left( {\frac{1}{2} - {q_0}} \right)
\end{equation}
This solution reproduces the SCM limit for which $\alpha  \to 0$. Indeed, in the SCM the cosmographic parameter $j$ is strictly equal to unity, and the parameter $C$ representing the equivalent of the relative density of dark energy is close to the observed value $C \approx 0.7$.
Let us now dwell on the question of the fulfillment of the second law of thermodynamics in the model under consideration \cite{31s}. As applied to the evolution of the late Universe, the answer to this question is simplified, since it is the entropy of the horizon that is the dominant component of the total entropy \cite{20s}.
Using the definition of the Bekenstein entropy, we represent it in the form
\begin{equation}\label{ehn107}
  {S_{BH}} = \frac{K}{{{H^2}}},\quad K = \frac{{\pi {k_B}{c^3}}}{{L_{Pl}^2}}
\end{equation}
Substituting the model solutions for the Hubble parameter (\ref{ehn99}), (\ref{ehn100}), we find
\begin{equation}\label{ehn108}
  {S_{BH}} = \frac{K}{{H_0^2}}{\left( {(1 - {C_\alpha }){{\left( {\frac{a}{{{a_0}}}} \right)}^{ - \frac{3}{2}(2 - \alpha )}} + {C_\alpha }} \right)^{\frac{2}{{\alpha  - 2}}}}\quad if\quad \alpha  \ne 2
\end{equation}
\begin{equation}\label{ehn109}
  {S_{BH}} = \frac{K}{{H_0^2}}{\left( {\frac{a}{{{a_0}}}} \right)^{3(1 - {C_\alpha })}}\quad if\quad \alpha  = 2
\end{equation}
Time derivatives of the Bekenstein entropy
\begin{equation}\label{ehn110}
  {\dot S_{BH}} =  - \frac{{2K\dot H}}{{{H^3}}} = 2{S_{BH}}\left( { - \frac{{\dot H}}{H}} \right)
\end{equation}
	
\begin{equation}\label{ehn111}
  {\ddot S_{BH}} = 2{S_{BH}}\left( {\frac{{3\dot H{H^2} - \ddot HH}}{{{H^2}}}} \right)
\end{equation}
Using solutions (\ref{ehn99}) and (\ref{ehn100}) we can study the behavior of the time derivatives of entropy as a function of the scale factor \cite{31s}.

In the model under consideration, both in the case $\alpha  \ne 2$ and in the case $\alpha  = 2$ the requirement $\frac{{\dot H}}{H} \le 0$  is fulfilled. Therefore, for the expanding Universe ($H > 0$) on the Hubble horizon the second law of thermodynamics is satisfied, namely ${\dot S_{BH}} \ge 0$.

The situation with  ${\ddot S_{BH}}$ is not so clear. We are interested in the asymptotic behavior of this derivative at $\frac{a}{{{a_0}}} \gg 1$. The analysis shows that the derivative ${\ddot S_{BH}}$ for $\alpha  < 2$ is positive at the early stage of evolution and negative at the last stage when approaching equilibrium. For $\alpha  > 2$ the condition  ${\ddot S_{BH}} < 0$ is not satisfied even in the asymptotics $\frac{a}{{{a_0}}} \gg 1$, which leads to violation of the second law of thermodynamics. Apparently, this is due to the fact that the values $\alpha  > 2$ are physically unacceptable. As we saw above (\ref{ehn106}) $\alpha  \sim (j - 1)$. In SCM the cosmographic parameter $j$  is strictly equal to unity. Values $\alpha  > 2$ requiring a large deviation $j$ from unity seem unlikely.

\subsection{Thermodynamics in the BV-model of entropic cosmology}

Let us now give an example of using the BV-model of entropic cosmology to describe a dissipative Universe with the birth of particles (matter). The adiabatic production of particles leads to the generation of irreversible entropy. The system of equations (\ref{eh10})-(\ref{eh12}) describes in the general case the evolution of the system in entropic cosmology. The term ${h_B}(t)$ in the considered model is associated with irreversible entropy due to the birth of particles. There may be other sources of irreversibility, such as bulk viscosity. Below we will assume that the Universe is dominated by non-relativistic matter and, therefore $p = 0(\omega  = 0)$. In this case, the background evolution is described by the equation
\begin{equation}\label{ehn112}
  \dot H =  - \frac{3}{2}{H^2} + \frac{3}{2}{f_\Lambda }(t) + {h_B}(t)
\end{equation}
It follows that the background evolution of the Universe in $L(t)$ and BV models is equivalent if the diriving terms are related  by ${h_B}(t) = \frac{3}{2}{f_\Lambda }(t)$. However, even in this case, the density perturbations in these models are different. For example, a constant term ${h_B}(t)$ leads to a non-zero term on the right side of the conservation equation, but a constant ${f_\Lambda }(t)$ does not.
Passing directly to the BV-model, we choose
\[{f_\Lambda }(t) = 0,\]
\begin{equation}\label{ehn113}
  {h_B}(t) = {C_{B\alpha }}H_0^2{\left( {\frac{H}{{{H_0}}}} \right)^\alpha }
\end{equation}
Here, as before, ${C_{B\alpha }}$ and $\alpha $ are dimensionless constants (real numbers). When the condition ${C_{B\alpha }} = \frac{3}{2}{C_\alpha }$ is satisfied, we can use formulas (\ref{ehn99}), (\ref{ehn100}) for the dependence of the Hubble parameter on the scale factor in the no- dissipative model. These two equations describe the background evolution of the Universe in the considered dissipative model with the driving term (\ref{ehn113}). However, in a dissipative Universe, irreversible entropy is produced due to the adiabatic creation of particles. To study the growth dynamics of irreversible entropy, it is necessary to relate the driving term ${h_B}(t)$ and the rate of 1 $\Gamma (t)$.
To do this, we turn to the fundamental system of equations for cosmological models with a variable number of particles \cite{34s},\cite{89s,90s,91s,92s}
\begin{equation}\label{ehn114}
  \dot n + 3Hn = \Gamma n
\end{equation}
\begin{equation}\label{ehn115}
  \dot s + 3Hs = \Gamma s
\end{equation}
\begin{equation}\label{ehn116}
  \dot \rho  + 3H\left( {\rho  + p + {p_c}} \right) = 0
\end{equation}
where $n,s,\rho $ are the particle number density, entropy density and energy density, a is the speed of creation of particles (matter), ${p_c}$ is the additional pressure generated by the process of the matter creation \cite{89s,90s,91s,92s}
\begin{equation}\label{ehn117}
  {p_c} =  - \frac{\Gamma }{{3H}}\left( {\rho  + p} \right)
\end{equation}
Comparing (\ref{eh12}) and (\ref{ehn106}) at ${f_\Lambda }(t) = 0$, we find the connection of interest to us
\begin{equation}\label{ehn118}
  \Gamma  = \frac{{3H{h_B}(t)}}{{4\pi G\rho }}
\end{equation}
Now we have all the necessary tools to calculate the evolution of entropy that interests us. Let's start by calculating the entropy of matter inside the Hubble volume. Normalized solution of equation (\ref{ehn115})
\begin{equation}\label{ehn119}
  \frac{s}{{{s_0}}} = \exp \left[ {\int_1^{\tilde a} {\left[ {\frac{{\Gamma \left( {\tilde a'} \right)}}{H} - 3} \right]} \frac{{d\tilde a'}}{{\tilde a'}}} \right]
\end{equation}
Here ${s_0}$ - the current value of the entropy density and $\tilde a \equiv \frac{a}{{{a_0}}}$. The evolution of the entropy density depends on two factors: the rate of matter creation and the rate of background evolution. Using connection (\ref{ehn118}), we find
\begin{equation}\label{ehn120}
  \frac{s}{{{s_0}}} = {\left( {\frac{H}{{{H_0}}}} \right)^2}
\end{equation}
Passing from the entropy density to the volume entropy, we get
\begin{equation}\label{ehn121}
  \frac{{{S_m}}}{{{S_{m0}}}} = \frac{{sr_H^3}}{{{s_0}r_{0H}^3}} = {\left( {\frac{H}{{{H_0}}}} \right)^2}{\left( {\frac{H}{{{H_0}}}} \right)^{ - 3}} = {\left( {\frac{H}{{{H_0}}}} \right)^{ - 1}}
\end{equation}
This expression is convenient to use to calculate the time derivatives of the entropy of matter. As for the entropy of the horizon, if the coincidence of the background evolution is taken into account and in the dissipative case, the results of the $\Lambda (t)$-model can be used.  Thus, we can construct two sets of functions  ${S_{BH}},{\dot S_{BH}},{\ddot S_{BH}}$ and  ${S_m},{\dot S_m},{\ddot S_m}$necessary for testing  second law of thermodynamics in the model under consideration. Analysis executed in [93] lead to the following results.

The BV-model of the dissipative Universe considered in this section always satisfies the second law of thermodynamics both for ${S_m}$ and for ${S_{BH}}$ , i.e. ${\dot S_m} \ge 0$ and ${\dot S_{BH}} \ge 0$. Consequently, the generalized second law of thermodynamics in the form ${\dot S_m} + {\dot S_{BH}} \ge 0$ is also realized for the BV-model with the driving term proportional to ${H^\alpha }$. The situation with entropy maximization seems to be more complicated. When $\alpha  < 2$, entropy maximization ${S_m}$ , i.e ${\ddot S_m} < 0$., is always performed. On the contrary, even if $\alpha  < 2$, the condition ${\ddot S_{BH}} < 0$ is satisfied only at a late stage of the evolution of the Universe, providing the final entropy maximization. Therefore, maximizing the total entropy depends almost entirely on the constraint on the horizon entropy

\section{Entropic limitations on the kinematics of models with the matter creation}

As an example of the effectiveness of entropic cosmology, let us examine the limitations that the laws of thermodynamics impose on the kinematics of cosmological expansion in the model of the matter creation. From the conditions on the first and second time derivatives of the total entropy, one can obtain important restrictions on various variants of the evolution of the Universe and the allowable intervals for varying the model parameters. The connection between thermodynamics and the description of the evolution of the Universe is well known and is embedded in the Friedman equations governing this evolution. Indeed, consider as an example the adiabatic expansion of the Universe filled with nonrelativistic matter $\left( {p = 0} \right)$  with density  $\rho $. In this case, the first law of thermodynamics $dE + pdV = 0$ is transformed into  $dE = d\left( {\rho V} \right) = d\left( {\rho {a^3}} \right) = 0$, where $a$ is scale factor.  Hence it follows that $\rho  \propto {a^{ - 3}}$ and from the first Friedmann equation $a \propto {t^{2/3}}$. Then the deceleration parameter is $q =  - \frac{{\ddot a}}{{a{H^2}}} = \frac{1}{2}$ . The last result means that the only variant of kinematics is compatible with the considered type of evolution of the Universe $\left( {dS = 0,p = 0} \right)$: decelerated expansion with the deceleration parameter $q = 1/2$ .
For models with the birth of nonrelativistic matter, the initial system of equations can be chosen in the form
\[{H^2} = \frac{{8\pi G}}{3}\rho ,\]
\begin{equation}\label{ehn122}
  \dot \rho  + 3H\rho  = \Gamma \rho
\end{equation}
or in terms of particle number density $n$
\[{H^2} = \frac{{8\pi m}}{{3m_P^2}}n,\]
\begin{equation}\label{ehn123}
  \dot n + 3Hn = \Gamma n
\end{equation}
Here $m$ is the mass of particles that form cold matter, $\Gamma$ is the rate of matter creation. Assuming that the rate of matter creation  is a function of the Hubble parameter $\Gamma  = \Gamma \left( H \right)$, it is easy to find an equation for finding of the Hubble parameter
\begin{equation}\label{ehn124}
  \dot H + \frac{3}{2}{H^2}\left( {1 - \frac{\Gamma }{{3H}}} \right) = 0
\end{equation}
We see that the dynamics of the model is determined by the competition between the rate of matter creation and the relative volume change. $\dot V/V = 3H$. Using the definition of the deceleration parameter $q =  - \frac{{\ddot a}}{{a{H^2}}}$ and the identity $\dot H + {H^2} = \frac{{\ddot a}}{a}$, we transform equation (124) to the form
\begin{equation}\label{ehn125}
  q =  - 1 + \frac{3}{2}\left( {1 - \frac{\Gamma }{{3H}}} \right)
\end{equation}
In the case $\Gamma  = 0$, relation (\ref{ehn125}) reproduces the cold matter deceleration parameter $q =  + 1/2$.
As before, we will assume that the total entropy of the system is the sum of the entropy of the horizon ${S_h}$ and the entropy of matter inside the horizon ${S_m}$. As the horizon entropy, we choose the Bekenstein-Hawking entropy ${S_h} = {S_{BH}} = \frac{{{k_B}\pi }}{{L_P^2{H^2}}}$. With regard to entropy ${S_m}$, it is natural to consider that each individual particle located inside the horizon contributes to the entropy equal to one bit, \cite{94s}. In this case
\begin{equation}\label{ehn126}
  {S_m} = {k_B}\frac{{m_P^2}}{{2Hm}}
\end{equation}
Therefore, the total entropy in the considered model
\begin{equation}\label{ehn127}
  S(t){\rm{ }} = {\rm{ }}{k_B}\left( {\frac{\pi }{{l_{Pl}^2H{{\left( t \right)}^2}}} + \frac{{m_P^2}}{{2H\left( t \right)m}}} \right)
\end{equation}
It is important to note that the total entropy (as well as the entropy of its components) depends only on the Hubble parameter, so the time derivatives of the entropy can be expressed in terms of the time derivatives of the Hubble parameter. The latter are connected with the cosmographic parameters with the help of relations,
\[\dot{H}=-H^2 (1+q)\]
\[\ddot{H}=H^3 (j+3q +2)\]
\[\dddot{H}=H^4 (s-4j -3q (q+4)-6)\]
\begin{equation}\label{ehn128}
  \ddddot{H}=H^5 (l-5s +10(q+2)j+30(q+2)q +24)
\end{equation}
We emphasize that this relationship has a model-free character.
The ratio  $\frac{{{S_h}}}{{{S_m}}}$ is proportional to the dimensionless parameter $\frac{m}{H}$.

In the early Universe (large values $H$, small horizon areas) the contribution of matter entropy dominates. As the Hubble parameter decreases, the dominating role passes to the entropy of the horizon.

\begin{figure}
  \centering
  \includegraphics[width=5 cm]{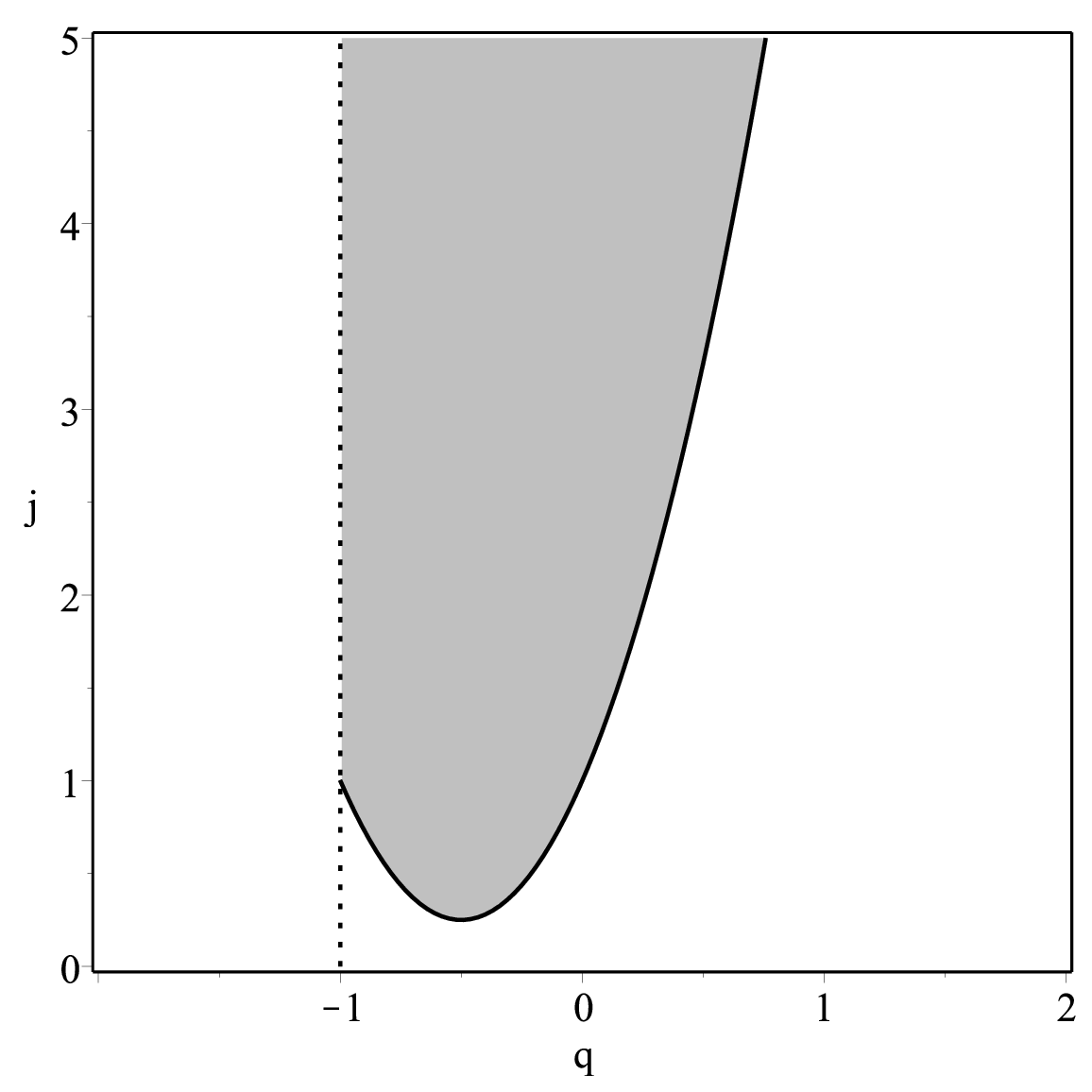}\\
  \caption{The domain of fulfillment of inequalities (130) and (133)) on the plane of cosmographic parameters   is shown in blue.}\label{fg1}
\end{figure}

Let us now proceed directly to finding the thermodynamic limitations imposed by the second law of thermodynamics on the kinematics of the expansion of the Universe. Differentiating with respect to time the expression for the total entropy (\ref{ehn127}) and using $\dot H =  - {H^2}(1 + q)$, we find
\begin{equation}\label{ehn129}
  {\rm{\dot S }} = {\rm{ }}\pi {k_B}\left( {1 + q} \right)\left( {\frac{2}{{l_{Pl}^2H\left( t \right)}} + \frac{{m_P^2}}{{2\pi m}}} \right)
\end{equation}
Whence it follows that the positivity of the first derivative of entropy in all models withe matter creation is satisfied under the condition
\begin{equation}\label{ehn130}
  q \ge  - 1
\end{equation}
Second time derivative of entropy
\begin{equation}\label{ehn131}
  \ddot S{\rm{  = }}{{\rm{k}}_B}\left( {{\rm{ }}\frac{{2\pi \left( {1 + q\left( t \right)} \right)}}{{L_{Pl}^2H{{\left( t \right)}^2}}}\frac{{dH(t)}}{{dt}} + \left( {\frac{{2\pi }}{{L_{Pl}^2H\left( t \right)}} + \frac{{m_P^2}}{{2m}}} \right)\frac{{dq\left( t \right)}}{{dt}}} \right)
\end{equation}
Using $\dot H =  - {H^2}(1 + q)$ and $\dot q =  - H\left( {j - 2{q^2} - q} \right)$ transform the condition $\ddot S{\rm{  < 0}}$ into
\begin{equation}\label{ehn132}
  \left( {3{q^2} - j + 3q + 1} \right) - \frac{{H\left( t \right)L_{Pl}^2{m_P}}}{{4\pi m}}\left( { - 2{q^2} + j - q} \right){\rm{ }} < {\rm{ }}0
\end{equation}
We are interested in the asymptotics of this inequality at $t \to \infty $ In this limit $\frac{H}{m} \ll 1$, and (\ref{ehn132}) is transformed into a simple inequality
\begin{equation}\label{ehn133}
  3{q^2} - j + 3q + 1 < 0
\end{equation}
The domain in which conditions (\ref{ehn130}) and (\ref{ehn133}) are satisfied is shown in Fig.\ref{fg1}
The obtained limitations on the cosmographic parameters are universal and must be satisfied for various models with the  of matter creation. Actually, the consistency of the proposed models with these universal limitations determines the permissible range of parameters included in these models.
We now test the limitations obtained above on the basis of a specific model with  matter creation.. Various variants of dependency $\Gamma (H)$ have been analyzed in [95 using a range of statistical criteria. Preference was given to a dependency with three free parameters $\alpha ,\beta ,n$
\begin{equation}\label{ehn134}
  \Gamma (H) = 3\beta H + 3\alpha {H_0}{\left( {\frac{{{H_0}}}{H}} \right)^n}
\end{equation}
In \cite{96s}, the thermodynamic limitations were formulated in terms of the free model parameters. In order to compare these results with those obtained above, it is necessary to express the model parameters in terms of cosmographic parameters. To do this, we use the procedure proposed in \cite{18s}.
Consider a simplified version: finding parameters $\alpha ,\beta $ for a fixed $n$. System of equations for determining parameters $\alpha ,\beta $
\[3\alpha {H_0}{\left( {\frac{{{H_0}}}{H}} \right)^n} + \left( { - 1 + 2q + 3\beta } \right)H = 0,\]
\begin{equation}\label{ehn135}
  - 3\alpha \left( {n - 2} \right)\left( {1 + q} \right){H_0}{\left( {\frac{{{H_0}}}{H}} \right)^n} + \left( { - 3 + 2j + q + 2{q^2} + 9\beta  + 9q\beta } \right)H = 0
\end{equation}
Solutions of this system
\begin{equation}\label{ehn136}
  \alpha  =  - \frac{{2\left( { - j + q + 2{q^2}} \right){{\left( {\frac{H}{{{H_0}}}} \right)}^n}H}}{{3\left( {1 + q} \right)\left( {1 + n} \right){H_0}}}
\end{equation}
\begin{equation}\label{ehn137}
  \beta  =  - \frac{{ - 1 + 2j - n - q + nq - 2{q^2} + 2n{q^2}}}{{3\left( {1 + q} \right)\left( {1 + n} \right)}}
\end{equation}
The verification of these solutions can be performed by passing to the limit  of a spatially flat Universe filled with nonrelativistic matter, with a zero rate of matter creation : $\Gamma  = 0 \to \alpha  = \beta  = 0$. In this case $q = 1/2,j = 1$. It is easy to check that
\[\alpha \left( {q = 1/2,j = 1} \right) = 0,\]
\begin{equation}\label{ehn138}
  \beta \left( {q = 1/2,j = 1} \right) = 0
\end{equation}
Since it is assumed that the parameters $\alpha ,\beta $ are time-independent constants, in relations (\ref{ehn136}), (\ref{ehn137}) one can use the current values of the cosmographic parameters
\[\alpha  =  - \frac{{2\left( { - {j_0} + {q_0} + 2q_0^2} \right)}}{{3\left( {1 + {q_0}} \right)\left( {1 + n} \right)}},\]
\begin{equation}\label{ehn139}
  \beta  =  - \frac{{ - 1 + 2{j_0} - n - {q_0} + n{q_0} - 2q_0^2 + 2nq_0^2}}{{3\left( {1 + {q_0}} \right)\left( {1 + n} \right)}}
\end{equation}
Of course, the verification of relations (\ref{ehn126}), (\ref{ehn127}) can also be performed by direct calculation of time derivatives $\dot \alpha ,\;\dot \beta $.

Using the obtained expressions for the parameters $\alpha ,\beta $, we can compare our general constraints (\ref{ehn130}), (\ref{ehn132}) with the constraints for a specific model $\Gamma (H)$. For the $\Gamma (H) = 3\beta H + 3\alpha {H_0}{\left( {\frac{{{H_0}}}{H}} \right)^n}$ condition $\dot S \ge 0$ is transformed into
\begin{equation}\label{ehn140}
  1 - \alpha  - \beta  \ge 0
\end{equation}
An important combination for analysis
\begin{equation}\label{ehn141}
  \alpha  + \beta  = \frac{1}{3}\left( {1 - 2{q_0}} \right)
\end{equation}
Therefore, the condition $\dot S \ge 0 \to 1 - \alpha  - \beta  \ge 0$ is transformed into a limitation for the deceleration parameter $q \ge  - 1$, which coincides with the constraint (\ref{ehn130}) obtained in the general case.

Avoiding cumbersome calculations, let us consider the kinematic limitations associated with the condition $\ddot S < 0$ for the particular case $\beta  = 0,n = 1$. In this case $j = 1$, for the parameter $\alpha $ we find
\begin{equation}\label{ehn142}
  \alpha  = \frac{1}{3}\left( {1 - 2q} \right)\frac{{{H^2}}}{{H_0^2}} \to \alpha  = \frac{1}{3}\left( {1 - 2{q_0}} \right)
\end{equation}
The fulfillment of fundamental inequalities $\dot S(t) \ge 0$, $\ddot S(t \to \infty ) < 0$ leads to the following interval of admissible values of the parameter $\alpha $
\begin{equation}\label{ehn143}
  0 < \alpha  < 1
\end{equation}
or in terms of the deceleration parameter
\begin{equation}\label{ehn144}
  - 1 < q < \frac{1}{2}
\end{equation}
Taking into account the condition  $\ddot S < 0$ leads to the appearance of an upper bound on the interval of admissible values of the deceleration  parameter.

\section{Entropic dark energy}

The idea of holographic dark energy \cite{97s,98s,99s,100s,101s,102s,103s,104s,105s} is extremely simple. The holographic formalism will lead to the appearance of an additional term in the Friedmann equation, which can be interpreted as dark energy, if this term generates negative pressure in the acceleration equation. Recall that holographic dynamics and entropic dynamics are almost synonymous. They are related by the central role of the surface, which carries all the information about the dynamics of the volume inside the surface. Above, referring to the need to take into account the contribution of surface terms in models with cosmological horizons, we manually introduced additional terms ${f_\Lambda }(H)$ and ${h_B}(H)$ into the Friedmann equation and (or) the acceleration equation, or only additional entropy pressure ${p_c}$ into the acceleration equation.  Below we consider an alternative approach based on the introduction of a new component in the energy budget -- holographic (entropic) dark energy. Based on firmly established thermodynamic relations (in particular, on the thermodynamics of black holes), we construct the energy density of this component and consider a number of examples.

In any effective quantum field theory defined over a spatial domain with a characteristic size $L$ and using the ultraviolet cutoff $\Lambda$, entropy is $S \sim {\Lambda ^3}{L^3}$. For example, fermions located at the sites of a spatial lattice of characteristic size $L$ withthe period ${\Lambda ^{ - 1}}$, are in one of ${2^{{{(\Lambda L)}^3}}}$ the states. Therefore, the entropy of such a system is $S \sim {\Lambda ^3}{L^3}$. We see that the entropy of a physical system depends on two values of fundamentally different scales: the macro scale $L$ and the micro scale ${\Lambda ^{ - 1}}$. In the example above, these are the size of the system and the lattice period. Therefore, any entropy restrictions lead to a connection between the macro and micro scales. According to the holographic principle, the upper limit of entropy is the Bekenstein-Hawking entropy
\begin{equation}\label{ehn145}
  {\Lambda ^3}{L^3} \le {S_{BH}} = \frac{A}{{4l_{Pl}^2}} = \pi {L^2}m_p^2
\end{equation}
Here ${S_{BH}}$ is the entropy of a black hole with a gravitational radius $L$. Inequality (\ref{ehn145}) is called the IR-UV (infrared-ultraviolet) correspondence \cite{106s}.
According to this inequality, the value of the infrared cutoff $L$ is strictly related to the value of the ultraviolet cutoff ${\Lambda ^{ - 1}}$. In other words, UV-scale physics depends on arameters IR-scale physics. In particular, in the case of saturation of the inequality (\ref{ehn145})
\begin{equation}\label{ehm146}
  L \sim {\Lambda ^{ - 3}}m_p^2
\end{equation}
The IR-UV correspondence can be reformulated as a constraint on the energy density ${\rho _\Lambda }$ in an arbitrary volume: the total energy contained in a region with linear size $L$ must not exceed the mass of a black hole of the same size, i.e.
\begin{equation}\label{ehn147}
  {L^3}{\rho _\Lambda } \le {M_{BH}} \sim Lm_p^2
\end{equation}
If this inequality is violated, a black hole is formed with an event horizon that prevents a further increase in the energy density. Let us now apply inequality (\ref{ehn147}) to the Universe as a whole. In this case, it is natural to identify the IR scale $L$  with the Hubble radius  ${H^{ - 1}}$, and to understand  ${\rho _\Lambda }$ as the density of the dominant component that fills the Universe, i.e. in the SCM  dark energy in the form of a cosmological constant. Then for the upper limit of the energy density we find
\begin{equation}\label{ehn148}
  {\rho _\Lambda } \sim {L^{ - 2}}m_p^2 \sim {H^2}m_p^2
\end{equation}
Given that
\[{m_{Pl}} \simeq 1.2 \times {10^{19}}{\kern 1pt} {\kern 1pt} GeV,\]
\begin{equation}\label{ehn149}
  {H_0} \simeq 1.6 \times {10^{ - 42}}{\kern 1pt} {\kern 1pt} GeV,
\end{equation}
we get
\begin{equation}\label{ehn150}
  {\rho _\Lambda } \approx {10^{ - 46}}{\kern 1pt} {\kern 1pt} Ge{V^4}
\end{equation}
This value is close (the difference is only two orders of magnitude and not 120) to the observed dark energy density $\rho _{DE}^{(obs)} \approx {10^{ - 48}}{\kern 1pt} {\kern 1pt} Ge{V^4}$.

The result seems extremely interesting, but its significance should not be exaggerated. It does not represent a solution to the problem of the cosmological constant, since we have not identified the nature of the phenomenologically introduced dark energy, but only an indication of the direction in which this solution should be sought. It is quite probable that entropy cosmology may turn out to be this direction.

Fundamental relation (\ref{ehn147}) can be rigorously reformulated in terms of horizon entropy when the latter is proportional to the area
\begin{equation}\label{ehn151}
  {\rho _\Lambda } \le \frac{S}{{{L^4}}}
\end{equation}
It is important to emphasize that for the horizon entropy, which is different from the Bekenstein entropy, relation (\ref{ehn152}) is only a hypothesis. Namely, the authors of \cite{106s} suggested that an effective field theory relating the energy density and the length scale through the saturation entropy could adequately describe the observed dark energy density for entropy other than the Bekenstein entropy. The validity of this hypothesis for generalized entropies can be confirmed (or refuted) by comparison with observational data.

\begin{figure}
  \centering
  \includegraphics[width=7 cm]{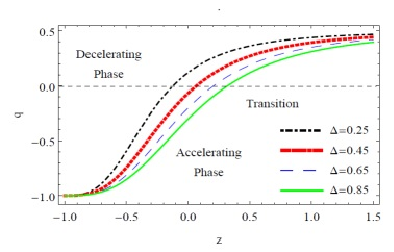}\\
  \caption{Plot of deceleration parameter $q$ with redshift $z$ \cite{107s}}\label{fg2}
\end{figure}

As an example, consider the holographic (entropy) dark energy in a flat FLRW Universe, constructed using the Barrow entropy \cite{59s}. The latter is a generalization of the Bekenstein entropy
\begin{equation}\label{ehn152}
  {S_B} = {\left( {\frac{A}{{{A_0}}}} \right)^{1 + \frac{\Delta }{2}}}
\end{equation}
where $A$ is the horizon area and  ${A_0}$ is the Planck area. Free model parameter $0 \le \Delta  \le 1$. The choice of entropy of this type is dictated by the desire to take into account the quantum deformations of the surface of the black hole event horizon. The measure of this deformation, which leads to the fractal structure of the horizon, is the new parameter $\Delta$. This one takes the value \[\Delta  = 0\] in the standard undeformed case of the Bekenstein entropy. The case $\Delta  = 1$ corresponds to the maximum deformation: an increase in the fractal dimension by one. Entropic dark energy density ${\rho _e}$ generated by entropy (\ref{ehn152})
\begin{equation}\label{ehn153}
  {\rho _e} = C{L^{\Delta  - 2}}
\end{equation}
where $C$  is the dimensional parameter $[C] = {L^{ - \Delta  - 2}}$ and $L$ is the infrared cutoff scale. At  $\Delta  = 0$, when the Barrow entropy reduces to the Bekenstein entropy, the entropic energy density ${\rho _e}$ reduces to the standard holographic energy density ${\rho _\Lambda } \sim {L^{ - 2}}$. Choosing the Hubble radius as ${H^{ - 1}}$ as the infrared macro scale, we get
\begin{equation}\label{ehn154}
  {\rho _e} = C{H^{2 - \Delta }}
\end{equation}
The evolution of a flat FLRW Universe filled with non-relativistic matter and entropic dark energy is described by a system of equations
\begin{equation}\label{ehn155}
  {H^2} = \frac{{8\pi G}}{3}({\rho _m} + {\rho _e})
\end{equation}
\begin{equation}\label{ehn156}
  {\dot \rho _m} + 3H{\rho _m} = 0
\end{equation}
\begin{equation}\label{ehn157}
  {\dot \rho _e} + 3H({\rho _e} + {p_e}) = 0
\end{equation}
Introducing the relative densities of matter ${\Omega _m} = \frac{{8\pi G}}{{3{H^2}}}{\rho _m}$ and dark energy, ${\Omega _e} = \frac{{8\pi G}}{{3{H^2}}}{\rho _e}$ equation (\ref{ehn155}) can be represented in the form
\begin{equation}\label{ehn158}
  {\Omega _m} + {\Omega _e} = 1
\end{equation}
Using the definition of entropy dark energy (\ref{ehn154}), we find \cite{107s}
\begin{equation}\label{ehn159}
  {\dot \rho _e} = \frac{{3C}}{2}(2 - \Delta ){H^{2 - \Delta }}\left( {\frac{{\Delta  \cdot {\Omega _e}}}{{(\Delta  - 2){\Omega _e} + 2}} - 1} \right)
\end{equation}
Combination
\begin{equation}\label{ehn160}
  \frac{{\dot H}}{{{H^2}}} = \frac{3}{2}\left( {\frac{{\Delta  \cdot {\Omega _e}}}{{(\Delta  - 2){\Omega _e} + 2}} - 1} \right)
\end{equation}
allows  to build the deceleration parameter
\begin{equation}\label{ehn161}
  q =  - 1 - \frac{{\dot H}}{{{H^2}}} = \frac{{1 - (\Delta  + 1){\Omega _e}}}{{(\Delta  - 2){\Omega _e} + 2}}
\end{equation}
Evolution $q$ as a function of redshift $z$ for different values of the parameter $\Delta $ is shown in Fig.\ref{fg2}. As can be seen from Fig.\ref{fg2}, the Barrow entropy dark energy model can well explain the history of the Universe, starting from the period of matter dominance ($q = 1/2$ ) and ending with the de Sitter asymptotics at $z \to  - 1$ ( $q =  - 1$) \cite{107s}.

For all considered values of the parameter, $\Delta$ there is a transition from the decelerated expansion to an accelerated one (change of the sign of the deceleration parameter). The value of the redshift at which this transition occurs can be consistent with the observations by selecting a parameter $\Delta $ in the interval $0 \le \Delta  \le 1$.

Thus, the first simple estimates show that the Barrow entropic  dark energy model is quite competitive. A broader analysis \cite{107s} confirms this conclusion.

\section{Conclusion}

This review presents the conceptual foundations of entropy cosmology, including both the already achieved results and emerging and still unsolved problems.

Entropic cosmology provides a fundamentally new scenario for the evolution of the Universe, generally consistent with observations, without the use of dark energy. Entropic pressure, like the negative pressure of dark energy, is responsible for the accelerated expansion of the Universe.
We have set as our goal, omitting the details that can be found in the references given, to present the conceptual foundations of entropic cosmology.  The key role in the considered approach is played by the thermodynamics of the horizon or, more specifically, its thermodynamic characteristics -- entropy and temperature. In entropic cosmology, the apparent horizon of the Universe is usually chosen as the horizon, which for a spatially flat Universe coincides with the Hubble horizon. We present arguments justifying this choice.

Much attention is paid to the problem of choosing the entropy of the cosmological horizon. By calculating the energy flux through the Hubble horizon, we demonstrate that the fulfillment of the Clausius relation $\delta Q = TdS$ and the standard Friedmann equations (i.e. GR) uniquely lead to the Bekenstein entropy. Due to the importance of this result, we also reproduced it within the framework of emergent cosmology. In essence, this approach can be treated as a microscopic view of entropic cosmology. As in the latter, evolution is determined by the interaction of volume and surface, but this dynamics is formulated in terms of microscopic degrees of freedom. However, so far this is only ''phenomenological microscopic'', since the nature of these degrees of freedom is unknown.

Two formulations of the basic equations of entropic cosmology are considered in detail. The first of them is based on the direct consideration of entropic forces (or, equivalently, the consideration of the contribution of surface terms when varying the Hilbert-Einstein action). Accounting for the action of entropic forces is achieved by modifying the Friedmann equation and the accelerated equation by introducing phenomenological deriving terms. Two types of modification are considered: for a non-dissipative Universe ($\Lambda (t)$-model) and a dissipative Universe (BV-model).
An alternative approach to formulating the basic equations of entropic cosmology is to treat the contribution of surface terms as holographic (entropic) dark energy. This allows us to keep the standard form of Friedmann equations. The equation of state and other characteristics of this type of dark energy depend on the choice of entropy on which it is based. As an example, we considered entropic dark energy constructed using Barrow entropy.


\begin{thebibliography}{31}

\bibitem{1s}S. Perlmutter, et al., Measurements of $\Omega$ and $\Lambda$ from 42 High-Redshift Supernovae,
Astrophys. J. 517, p.565-586 (1999)

\bibitem{2s} A. G. Ress, et al., Observational Evidence from Supernovae for an Accelerating Universe and a Cosmological Constant, Astron. J. 116, p.1009-1038, (1998)

\bibitem{3s}D. A. Easson, P. H. Frampton, and G. F. Smoot, Entropic Accelerating Universe,
 Phys. Lett. B 696, p.273-277, (2011)

\bibitem{4s}D. A. Easson, P. H. Frampton, and G. F. Smoot,, Entropic inflation, Int. J. Mod. Phys. A 27, 1250066 (2012)

\bibitem{5s}Bekenstein, J. D., Black holes and the second law, Lett. Nuovo Cimento Soc. Ital. Fis. 4, p.737-740, (1972)

\bibitem{6s}Bekenstein, J. D. Black holes and entropy, Phys. Rev. D 7, p.2333-2346,  (1973)

\bibitem{7s}Bekenstein, J. D. Generalized second law of thermodynamics in Black hole physics, Phys. Rev. D 9, p.3292--3300, (1974)

\bibitem{8s} Bekenstein, J. D., Statistical Black Hole Thermodynamics, Phys. Rev. D 12, p.3077-3085, (1975)

\bibitem{9s}Bardeen, J.M., B.Carter and S.W.Hawking, The four laws of black hole mechanics, Commun. Math.Phys.31, p.161-170,(1973)

\bibitem{10s} S.W. Hawking, Particle Creation by Black Holes, Commun. Math. Phys. 43,  p.199-220, (1975)

\bibitem{11s} S.W. Hawking, Black Holes and Thermodynamics, Phys. Rev. D 13, p.191--197, (1976)

\bibitem{12s} T. Jacobson, Thermodynamics of space-time: The Einstein equation of state, Phys. Rev. Lett. 75, p.1260-1263, (1995)

\bibitem{13s} W. Rindler, Relativity, Published in the United States by Oxford University Press Inc., New York (2006)

\bibitem{14s}W. Rindler,Hyperbolic motion in curved space-time, Phys. Rev. 119, p.2082-2089 (1960)

\bibitem{15s}G.W. Gibbon and S.W.Hawking, Cosmological event horizons, thermodynamics, and particle creation,  Phys Rev D 15 p.2738-2751, (1977)

\bibitem{16s}Unruh, W., Notes on Black-Hole Evaporation, Phys. Rev. D 14, p.870-892, (1976)

\bibitem{17s}C.W. Misner and D.H. Sharp, Relativistic Equations for Adiabatic Spherically Symmetric Gravitational Collapse, Phys. Rev. 136,  p.571-576, (1964)

\bibitem{18s} W.C. Hernandez and C.W. Misner, Observer Time as a Coordinate in Relativistic Spherical Hydrodynamics, Astrophys. J. 143, p.452-464, (1966)

\bibitem{19s} V. Faraoni, Cosmological apparent and trapping horizons, Phys.Rev. D 84, 024003, (2011)

\bibitem{20s} Chas A. Egan, Charles H. Lineweaver, A Larger Estimate of the Entropy of the Universe, Astrophys.J.710:1825-1834, (2010)

\bibitem{21s}E.Verlinde, On the Origin of Gravity and the Laws of Newton,     JHEP 1104:029, (2011)

\bibitem{22s} Yu. L. Bolotin, A. V. Tur, V. V. Yanovsky, Physics of Limit Values at Planck scale, arXiv:2005.03984

\bibitem{23s} S.W. Hawking and G. T. Horowitz, The Gravitational Hamiltonian, Action, Entropy, and Surface TermsClass. Quant. Grav. 13, 1487 (1996)

\bibitem{24s} G. 't Hooft, Dimensional Reduction in Quantum Gravity, in Salamfestschrift, Editors: A. Ali, J. Ellis and S. Randjbar-Daemi, World Scientific Publishing Company (1994) p. 284, arXiv:gr-qc/9310026

\bibitem{25s} L. Susskind, The World as a Hologram, J. Math. Phys. 36, 6377 (1995), hep-th/9409089

\bibitem{26s} N. Komatsu and S. Kimura, Non-adiabatic-like accelerated expansion of the late universe in entropic cosmology, Phys. Rev. D 87, 043531 (2013)  arXiv:1307.5949v2 [astro-ph.CO

\bibitem{27s} N. Komatsu and S. Kimura, General form of entropy on the horizon of the universe in entropic cosmology, Phys. Rev. D 93, 043530 (2016), arXiv:1511.04364 [gr-qc]

\bibitem{28s} N. Komatsu and S. Kimura, Phys. Rev. D 90, 123516 (2014), arXiv:1408.4836 [astro-ph.CO]

\bibitem{29s} N. Komatsu and S. Kimura, Entropic cosmology in a issipative universe, Phys. Rev. D 89, 123501 (2014), arXiv:1402.3755 [astro-ph.CO]

\bibitem{30s} N. Komatsu, S. Kimura,  Evolution of the universe in entropic cosmologies via different formulations, Phys. Rev. D 89, 123501 (2014), arXiv:1402.3755v3 (astro-ph)

\bibitem{31s} N. Komatsu, Horizon thermodynamics in holographic cosmological models with a power-law term, Phys. Rev. D 100, 123545 (2019), arXiv:1911.08306v2 (gr-qc)

\bibitem{32s} N. Komatsu, Generalized thermodynamic constraints on holographic-principle-based cosmological scenarios,  Phys. Rev. D 99, 043523 (2019), arXiv:1810.11138v3 (gr-qc)

\bibitem{33s} J. D. Barrow, T. Clifton, Cosmologies with energy exchange, Phys. Rev. D 73, 103520(6), (2006)

\bibitem{34s} D. Kondepudi and I. Prigogine, Modern Thermodynamics: From Heat Engines to Dissipative Structures (John Wiley and Sons, New York, 1998)

\bibitem{35s} S. Weinberg, Gravitation and Cosmology (John Wiley and Sons, New York, 1972) 	

\bibitem{36s} J. D. Barrow, The deflationary universe: An instability of the de Sitter universe, Phys. Lett. B 180, p.335-339, (1986)

\bibitem{37s} W. Zimdahl, Bulk viscous cosmology, Phys. Rev. D 53, p.5483--5493, (1996)

\bibitem{38s} Brevik,  Cardy-Verlinde entropy formula in the presence of a general
cosmological state equation, Phys. Rev. D 65, 127302(4), (2002)

\bibitem{39s} S. Nojiri, S.D. Odintsov, Inhomogeneous equation of state of the universe: Phantom era, future singularity, and crossing the phantom barrier, Phys. Rev. D 72, 023003(12) (2005)

\bibitem{40s} N. Komatsu, S. Kimura, Cosmic microwave background radiation temperature in a dissipative universe, Phys. Rev. D 92, 043507(12), (2015)

\bibitem{41s}Y. B. Zel'dovich, Particle production in cosmology, Soviet Journal of Experimental and Theoretical Physics Letters 12, p.307-311, (1970)

\bibitem{42s} Fei-Quan Tu, Yi-Xin Chen, Qi-Hong Huang,A cosmic accelerated scenario based on degrees of freedom of the space-time, arXiv:2107.10682 (gr-qc)

\bibitem{43s} Fei-Quan Tu, Yi-Xin Chenb and Qi-Hong Huanga Thermodynamics in the universe described by the emergence of the space and the energy balance relation, arXiv:1808.10562

\bibitem{44s} T. Padmanabhan, Emergence and Expansion of Cosmic Space as due to the Quest for HolographicEquipartition, arXiv: 1206.4916

\bibitem{45s} T. Padmanabhan, Thermodynamical Aspects of Gravity: New insights, Rept. Prog. Phys. 73, 046901(10), (2010) 0911.5004

\bibitem{46s} A. Komar, Covariant Conservation Laws in General Relativity, Phys. Rev. 113, p.934-936, (1959)

\bibitem{47s} A. Komar, Positive-Definite Energy Density and Global Consequences for General Relativity, Phys.Rev. 129, p.1873-1876, (1963)

\bibitem{48s} T. Padmanabhan, Emergence and Expansion of Cosmic Space as due to the Quest for Holographic
Equipartition, arXiv, arXiv:1206.4916 (2012)

\bibitem{49s} R.G. Cai, First law of thermodynamics and Friedmann equations of Friedmann-Robertson-Walker universe, J. High Energy Phys., 16, p.1-12, (2012).

 \bibitem{50s} Fu-Wen Shu, Y. Gong, Equipartition of energy and the first law of thermodynamics at the apparent horizon, Int. J. Mod. Phys. D 20, p.553-559, (2011)

\bibitem{51s}J. P. Mimoso, D. Pavon, Fluctuations of the flux of energy on the apparent horizon, Phys. Rev. D 97, 103537(4), (2018)

\bibitem{52s} T. Padmanabhan, Equipartition of energy in the horizon degrees of freedom and the emergence of gravity, Mod. Phys. Lett. A 25, p.1129-1136, (2010)

\bibitem{53s} N. Komatsu, Energy stored on a cosmological horizon and its thermodynamic fluctuations in holographic equipartition law. Phys. Rev. D 105, 043534(13), (2022)

\bibitem{54s}  T. Padmanabhan, Entropy of Static Spacetimes and Microscopic Density of States, Class. Quantum Grav. 21, p.4485-4494, (2004)

\bibitem{55s} P. B. Krishna, T. K. Mathew, Entropy maximization in the emergent gravity paradigm, Phys. Rev. D 99, 023535(11), (2019)

\bibitem{56s} C. Tsallis, Possible generalization of Boltzmann-Gibbs statistics, J. Stat. Phys. 52, p.479-487, (1988)

\bibitem{57s} C. Tsallis and L. J. L. Cirto, Black hole thermodynamical entropy, Eur. Phys. J. C 73, 2487(7), (2013)

\bibitem{58s} A. Renyi (1961). On measures of information and entropy. Proceedings of the 4th Berkeley Symposium on Mathematics, Statistics and Probability 1960. pp. 547-561

\bibitem{59s} J. D. Barrow, The Area of a Rough Black Hole, Phys. Lett. B 808, 135643, (2020)

\bibitem{60s} S. Nojiri, S. D. Odintsov, T. Pau, Early and late universe holographic cosmology from a new generalized entropy, arXiv:2205.08876v1 [gr-qc]

\bibitem{60ss}S. D. Odintsov, T. Paul, Generalised (non-singular) entropy functions with applications to cosmology and black holes, arXiv:2301.01013v1 [gr-qc]

\bibitem{61s}M. Akbar and Rong-Gen Cai, Friedmann Equations of FRW Universe in Scalar-tensor Gravity,   Gravity and First Law of Thermodynamics Phys.Lett.B635 p.7-10, (2006)

\bibitem{62s} Rong-Gen Cai, Sang Pyo Kim, First Law of Thermodynamics and Friedmann Equations of Friedmann-Robertson-Walker Universe, JHEP 0502 (2005) 050

\bibitem{63s}D. Bak and S. J. Rey, Class., Cosmic Holography, Quant. Grav. 17, L83 (2000) [arXiv:hep-th 9902173]

\bibitem{64s}S. A. Hayward, S. Mukohyama and M. C. Ashworth, Dynamic black-hole entropy, Phys. Lett. A 256, 347 (1999) [arXiv:gr-qc 9810006]

\bibitem{65s}S. A. Hayward, Unified first law of black-hole dynamics and relativistic thermodynamics, Class. Quant. Grav. 15, p.3147-3162, (1998)[arXiv:gr-qc 9710089]

\bibitem{66s}Hareesh T, P. B. Krishna, and Titus K. Mathew, First Law of Thermodynamics and Emergence of Cosmic Space in a Non-Flat Universe, arXiv:1908.03349v2 [gr-qc]

\bibitem{67s}Hassan Basari V. T., P. B. Krishna, and Titus K. Mathew. Unified formalism for the emergence of space from the first law of Thermodynamic arXiv:2209.00304v1 [gr-qc]

\bibitem{68s} P. B. Krishna and Titus K. Mathew, Emergence of cosmic space and the maximization of horizon entropy, arXiv:2002.02121v2 (gr-qc)

\bibitem{69s} V.A. Antonov, Vest. leningr. gos. Univ. 7, 135 (1962)

\bibitem{70s} D. Lynden-Bell and R. Wood, The Gravo-Thermal Catastrophe in Isothermal Spheres and the Onset of Red-Giant Structure for Stellar Systems,  Mon. Not. R. Astron. Soc. 138, p.495-525, (1968)

\bibitem{71s} K.S. Thorne, W.H. Zurek, and R.H. Price, The thermal  atmosphere of a black hole, in Black Holes: The Membrana Paradigm, edited by K.S. Thorne, R.H. Price and D.A. Macdonald (Yale University Press, New Haven, (1986)

\bibitem{72s} Diego Pav\'{o}n and Ninfa Radicella, Does the entropy of the Universe tend to a maximum? arXiv:1209.3004v1 (gr-qc)

\bibitem{73s} H. Moradpour, A. H. Ziaie1y, Iarley P. Lobo, J. P. Morais Gra, U. K. Sharma, A. Sayahian Jahromi, The third law of thermodynamics and black holes, arXiv:2106.00378v1 [gr-qc]

\bibitem{74s} I.Racz, Class. Does the third law of black hole thermodynamics really have a serious failure? Quantum Grav. 17, p.4353-4356, (2000)

\bibitem{75s} D. Chen et al, Effects of quantum gravity on black holes, Int. J. Mod. Phys. A 29, 1430054 (2014)

\bibitem{76s}W. Xu, J. Wang, X. h. Meng, Thermodynamic relations for the entropy and temperature of multi-horizon black holes, Galaxies 3(1), 53-71, (2015)

\bibitem{77s}Y. Yao, M. S. Hou, Y. C. Ong, A complementary third law for black hole thermodynamics, Eur. Phys. J. C 79, 513 (2019)

\bibitem{78s} Swastik Bhattacharya and S. Shankaranarayanan, Is there an upper bound on the size of a black-hole? Int. J, Mod. Phys. D. 27, 1847011 (2018), arXiv:1805.05686v1 (gr-qc)

\bibitem{79s} Swastik Bhattacharya and S. Shankaranarayanan, Is there an upper bound on the size of a black-hole? Int. J, Mod. Phys. D. 27, 1847011 (2018), arXiv:1805.05686v1 (gr-qc)

\bibitem{80s}D. J. Zamora, C. Tsallis, Thermodynamically consistent entropic-force cosmology, arXiv:2201.01835v1 [gr-qc]

\bibitem{81s}D. J. Zamora, C. Tsallis, Thermodynamically consistent entropic ination including subdominant contribution, arXiv:2201.03385v1 [gr-qc]

\bibitem{82s} N. Komatsu, Thermodynamic constraints on a varying cosmological-constant-like term from the holographic equipartition law with a power-law corrected entropy, Phys. Rev. D 96, 103507 (2017)

\bibitem{83s}Yu.L. Bolotin, V.A. Cherkaskiy, O.A. Lemets, O.Yu. Ivashtenko, M.I. Konchatnyi, L.G. Zazunov, Applied cosmography: A Pedagogical Review, arXiv:1812.02394v1 [gr-q] (2018)

\bibitem{89s} M. O. Calv\={a}o, J. A. S. Lima, I. Waga, On the thermodynamics of matter creation in cosmology, Phys. Lett. A 162, p.223-226, (1992)

\bibitem{90s} J. A. S. Lima, A. S. M. Germano, On the equivalence of bulk viscosity and matter creation, Phys.Lett. A 170, p.373-378, (1992)

\bibitem{91s} J. A. S. Lima, R. C. Santos, J.V. Cunha, Is $\Lambda CDM$ an effective CCDM cosmology? J. Cosmol. Astropart. Phys. 03 p.027-(2016)

\bibitem{92s}J. A. S. Lima, I. Baranov, Gravitationally Induced Particle Production: Thermodynamics and Kinetic Theory, Phys. Rev. D 90, 043515, (2014)

\bibitem{93s} N. Komatsu, Entropy production due to adiabatic particle creation in a holographic dissipative cosmology, Phys. Rev. D 102, 063512 (2020)

\bibitem{94s} Mimoso, J. P., Pavon, D.,  Entropy evolution of universes with initial and final de Sitter eras, Phys. Rev. D , 87, 047302 (2013)

\bibitem{95s} J.F. Jesus, R. Valentim, F. Andrade-Oliveira, Thermodynamic constraints on matter creation models, J. Cosm. Astr. Phys., 09 1709 (2017) arXiv:1612.04077 [astroph. CO]

\bibitem{96s} R. Valentim, J. F. Jesus, Thermodynamic constraints on matter creation models, arXiv:1904.10313 [gr-qc]

\bibitem{97s} S. Wang, Y.Wang, M. Li, Holographic dark energy, Phys. Rept. 696, (2017)

\bibitem{98s} M. Li, A model of holographic dark energy, Phys. Lett. B 603, (2004)

\bibitem{99s} D. Pavon, W. Zimdahl, Holographic dark energy and cosmic coincidence, Phys. Lett. B 628, 206, (2005)

\bibitem{100s} E. Elizalde, S. Nojiri, S. D. Odintsov, P. Wang, Dark energy: vacuum fluctuations, the effective phantom phase, and holography, Phys. Rev. D 71, (2005) 103504
\bibitem{101s}  M. Li, C. Lin, Y. Wang, Some issues concerning holographic dark energy, JCAP 0805, 023, (2008)

\bibitem{102s} Y.G. Gong, B. Wang, Y.Z. Zhang, Holographic dark energy reexamined, Phys. Rev. D 72, 043510, (2005)

\bibitem{103s} Y. Gong, T. Li, A modified holographic dark energy model with infrared infinite extra dimension (s), Phys. Lett. B 683, 241, (2010)

\bibitem{104s} M. Malekjani, Generalized holographic dark energy model in the Hubble length, Astrophys. Space Sci. 347, 405, (2013)

\bibitem{105s}R. C. G. Landim, Holographic dark energy from minimal supergravity, Int. J. Mod. Phys.D 25, (2016) 1650050.12

\bibitem{106s}  A.G. Cohen, D.B. Kaplan and A.E. Nelson, Effective field theory, black holes, and the cosmological constant, Phys. Rev. Lett. 82 (1999) 4971

\bibitem{107s} Anirudh Pradhan, Archana Dixit, Vinod Kumar Bhardwaj, Barrow HDE model for Statefinder diagnostic in FLRW Universe, arXiv:2101.00176v1 [gr-qc]

\end{thebibliography}
\end{document}